\documentclass[preprint,linenumber,superscriptaddress,amsmath,amsfonts,nofootinbib]{revtex4-1}
\usepackage[pdftex]{color,graphicx}
\usepackage{amsthm,amsmath,amssymb}
\usepackage{mathrsfs}
\usepackage[colorlinks,linkcolor=blue,citecolor=blue,bookmarks,pdfstartview=FitH]{hyperref}
\usepackage{graphics}
\usepackage{epstopdf}
\usepackage{xcolor}
\usepackage[bottom]{footmisc}
\usepackage{colortbl,booktabs}
\usepackage{tabularx}
\makeatletter
\def\tagform@#1{\maketag@@@{\ignorespaces#1\unskip\@@italiccorr}}
\let\orgtheequation\theequation
\def\theequation{(\orgtheequation)}
\makeatother

\begin{document}
\title{Low-frequency shear Alfv\'en waves at DIII-D: theoretical interpretation of experimental observations}
\author{Ruirui Ma}
\email[corresponding author. Email address: ]{rrma@swip.ac.cn}
\affiliation{{\rm Southwestern Institute of Physics, P.O. Box 432, Chengdu, 610041, China}}
\affiliation{{\rm Center for Nonlinear Plasma Science and C.R. ENEA Frascati, C.P. 65, 00044 Frascati, Italy}}
\author{W.W. Heidbrink}
\affiliation{{\rm Department of Physics and Astronomy, University of California, Irvine, CA 92697-4574, USA}}
\author{Liu Chen}
\affiliation{{\rm Institute for Fusion Theory and Simulation and Department of Physics, Zhejiang University,
Hangzhou, 310027, People's Republic of China}}
\affiliation{{\rm Department of Physics and Astronomy, University of California, Irvine, CA 92697-4574, USA}}
\affiliation{{\rm Center for Nonlinear Plasma Science and C.R. ENEA Frascati, C.P. 65, 00044 Frascati, Italy}}
\author{Fulvio Zonca}
\affiliation{{\rm Center for Nonlinear Plasma Science and C.R. ENEA Frascati, C.P. 65, 00044 Frascati, Italy}}
\affiliation{{\rm Institute for Fusion Theory and Simulation and Department of Physics, Zhejiang University,
Hangzhou, 310027, People's Republic of China}}
\author{Zhiyong Qiu}
\affiliation{{\rm Institute for Fusion Theory and Simulation and Department of Physics, Zhejiang University,
Hangzhou, 310027, People's Republic of China}}
\affiliation{{\rm Center for Nonlinear Plasma Science and C.R. ENEA Frascati, C.P. 65, 00044 Frascati, Italy}}
\date{\today}
\begin{abstract}
  The linear properties of the low-frequency shear Alfv\'en waves such as those associated with the beta-induced Alfv\'en eigenmodes (BAEs) and the low-frequency modes observed in reversed-magnetic-shear DIII-D discharges (W. Heidbrink, {\it et al} 2021 {\it Nucl. Fusion} {\bf 61} 066031) are theoretically investigated and delineated based on the theoretical framework of the general fishbone-like dispersion relation (GFLDR). By adopting representative experimental equilibrium profiles, it is found that the low-frequency modes and BAEs are, respectively, the reactive-type and dissipative-type unstable modes with dominant Alfv\'enic polarization, thus the former being more precisely called low-frequency Alfv\'en modes (LFAMs). More specifically, due to different instability mechanisms, the maximal drive of BAEs occurs, in comparison to LFAMs, when the minimum of the safety factor ($q_{min}$) deviates from a rational number. Meanwhile, the BAE eigenfunction peaks at the radial position of the maximum energetic particle pressure gradient, resulting in a large deviation from the $q_{min}$ surface. Moreover, the ascending frequency spectrum patterns of the experimentally observed BAEs and LFAMs can be theoretically reproduced by varying $q_{min}$ and also be well interpreted based on the GFLDR. The present analysis illustrates the solid predictive capability of the GFLDR and its practical usefulness in enhancing the interpretative capability of both experimental and numerical simulation results.
\end{abstract}

\maketitle

\section{Introduction and Motivation}\label{introduction}
The low-frequency Alfv\'en wave spectrum in the kinetic thermal-ion (KTI) gap frequency range \cite{Chen2007} has been of research interest since the first observations of beta-induced Alfv\'en eigenmodes (BAEs) \cite{Heidbrink1993,Turnbull1993}. These modes are characterized with frequencies comparable to thermal ion transit and/or bounce frequencies, and can interact with both thermal and fast particles \cite{Zonca1996,Zonca1999,Zonca2010,Chavdarovski2009,Lauber2009,Chen2016}, with possible (positive/negative) impact on the corresponding transport processes resulting from finite fluctuation and zonal field structures levels \cite{Chen2007,Chen2016,Zonca2021a}. The effects of energetic particles (EPs) on low-frequency shear Alfv\'en waves (SAWs) ranging from kinetic ballooning mode (KBM) \cite{Cheng1982,Tang1980,Biglari1991} to BAE are one of areas widely studied in the magnetic fusion literature. Recent papers on this topic cover the interpretation and modeling of experimental measurements by currently developed innovative diagnostics \cite{Sharapov2013,Gorelenkov2014,Heidbrink2021,Heidbrink2021a,Heidbrink2021b}, as well as latest progress in comparing numerical investigation and/or simulation results with observed phenomena \cite{Curran2012,Lauber2013a,Chavdarovski2014,Fasoli2016,Bierwage2017,Choi2021a}.

A series of dedicated experiments have been recently conducted on DIII-D to investigate the stability of the low-frequency SAWs \cite{Heidbrink2021,Heidbrink2021a,Heidbrink2021b}. The experiments show that the observed low-frequency mode\footnote{We will refer from now on only to the low frequency Alfv\'en mode (LFAM) which belongs to low-frequency SAWs predominantly Alfv\'enic polarization, keeping in mind that this terminology is the same as the low-frequency mode observed in recent DIII-D experiments \cite{Heidbrink2021}.}, which was previously misidentified as `beta-induced Alfv\'en acoustic eigenmode (BAAE)' \cite{Gorelenkov2007,Gorelenkov2009}, is actually a lower-frequency reactive unstable KBM which favors high thermal electron temperature but almost has no coupling with energetic ions \cite{Heidbrink2021}; while the BAE is resonantly excited by energetic ions with its stability depending sensitively on the beam power and injection geometry \cite{Heidbrink2021a}, consistent with earlier theoretical predictions \cite{Chen2017} based on the GFLDR theoretical framework \cite{Zonca2014,Zonca2014a}. These instabilities are also found to occur when the minimum of the safety factor ($q_{min}$) approaches rational values and the modes in ascending pattern of higher frequency BAEs and LFAMs are separated by approximately the toroidal rotation frequency ($f_{rot}$). However, the subtle differences between them are that, for LFAMs, the maximum frequency appears at rational values of $q_{min}$ and the detected modes are radially localized near $q_{min}$, while BAEs occur at times near rational $q_{min}$ values but the timing of unstable modes is less precise than that for LFAMs. In addition, compared with the LFAMs, the BAE eigenfunction shows more deviation from the radial position of $q_{min}$ spatially. Although dedicated numerical simulations of the linear properties of the BAEs and LFAMs \cite{Varela2018,Choi2021a} have been carried out, the above experimental phenomena have not been fully explained. Motivated by this, the present work aims to provide an in-depth theoretical understanding of the linear properties of low-frequency SAWs, with particular attention to the effects of energetic ions on their stability. The analysis is carried out based on the theoretical framework of the generalized fishbone-like dispersion relation (GFLDR) \cite{Chen1984,Chen1994,Tsai1993,Zonca2006,Zonca2007,Zonca2014,Zonca2014a}, and provides qualitative and quantitative interpretation of the main instability mechanisms underlying the numerical simulation results and experimental observations. As a result, our analysis provides yet another evidence of the predictive strength of the GFLDR theoretical framework and of its enhanced ``interpretative capability for both experimental and numerical simulation results" \cite{Zonca2014,Zonca2014a}.

\begin{figure}[htbp]
\centering
\includegraphics[width=0.5\textwidth]{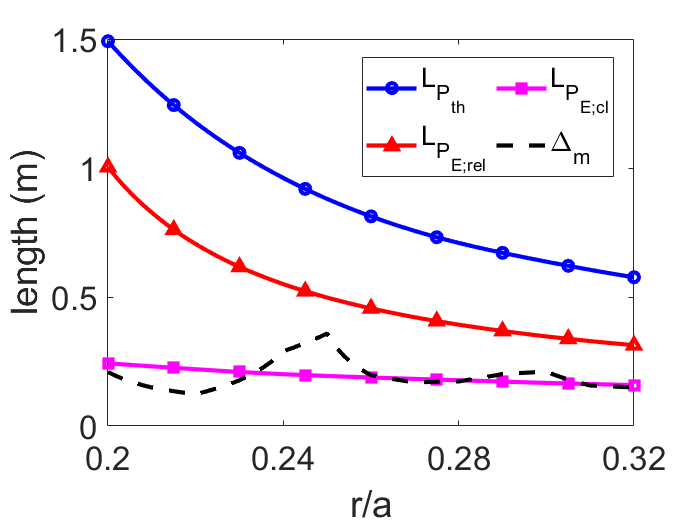}
\caption{The radial dependences of the typical scale lengths of thermal and energetic particle pressure ($L_{P_{th}}$ and $L_{P_{E}}$), as well as the estimated radial mode width ($\Delta_m$). }
\label{LP_scales}
\end{figure}

In this work, unlike the previous paper not considering effects due to energetic particles (EPs) \cite{Ma2022}, we focus on the BAE excitation via transit resonance with passing fast ions created by NBI heating \cite{Heidbrink2021a}. In this case, the dynamics of various species enter the dispersion relation of low-frequency SAW, and affect its behavior linearly at different pressure gradient scale lengths. For DIII-D discharge \#178631, Fig. \ref{LP_scales} shows the radial dependence of different scale lengths of thermal and energetic particle pressure ($L_{P_{th}}$ and $L_{P_{E}}$), as well as the estimated radial mode width ($\Delta_m$) for weak and/or vanishing magnetic shear range, i.e., $|s|=|(r/q)(dq/dr)|\lesssim0.05$. More specifically, the EP pressure profiles are given by the following two limits. One is the relaxed EP profile provided with EFIT reconstruction \cite{Lao1985}, where the fast-ion pressure is the difference between the equilibrium pressure and the thermal pressure. The other is the ``classical'' EP profile obtained by TRANSP/NUBEAM \cite{Pankin2004} in the absence of fast-ion transport by instabilities. The pressure scale lengths of EPs are denoted by $L_{P_{E;rel}}$ and $L_{P_{E;cl}}$ for these two cases (respectively). The true EP profile when the modes are destabilized likely lies between these two limits. The actual pressure is closest to the EFIT-based one but this is measured after the unstable modes have (presumably) caused the gradients to flatten. Meanwhile, for the weak and/or vanishing magnetic shear region and given toroidal and poloidal mode numbers $(n,m)$, the normalized parallel wave vector is $\Omega_{A,m}=k_{\parallel n0}q_{min}R_0=nq_{min}-m$, and the radial width of the mode can then be estimated by $\Delta_m\simeq 1/|nq''|^{1/2}$ \cite{Zonca2000,Zonca2002}. Here, $k_{\parallel n0}$ represents the parallel wave-vector at $r_0$, where $q$ has a minimum given by $q_{min}$, $q''$ denotes the second derivative of $q$ in the radial direction, and $R_0$ is the torus major radius.  It can be found that in this region, $L_{P_{th}}\gg \Delta_m$, which yields the usual local limit of the mode dispersion relation. This is the case for the reactive unstable LFAM in the absence of EPs already studied in Ref. \cite{Ma2022}. However, for the energetic ion-driven BAEs, there are two distinct cases: the moderate EP pressure gradient case with $L_{P_{E;rel}}>\Delta_m$, which also approximately yields the usual local GFLDR \cite{Tsai1993,Chen1994,Zonca1996,Zonca2000,Zonca2002,Zonca2007,Zonca2014,Zonca2014a}; and the strong EP pressure gradient case with $L_{P_{E;rel}}\simeq\Delta_m$, for which the global dispersion relation of low-frequency SAWs is needed and will be discussed in Sec. \ref{Theoretical Model}.
Performing detailed numerical investigations of the two cases, it is found that the LFAMs and BAEs can both be driven unstable, however, due to different instability mechanisms, these modes yield different experimental observations. All these features can be, quantitatively and qualitatively, interpreted theoretically based on the GFLDR. Moreover, it is also confirmed that the stability of BAAE is not affected by EPs, even though it becomes weakly damped after coupling with KBM, consistent with theoretical predictions by Chen and Zonca \cite{Chen2017} as well as numerical simulation results reported in Refs. \cite{Choi2021a,Lauber2013a,Bierwage2017}.

The paper is structured as follows. Local and global dispersion relations for the low-frequency SAWs near weak and/or vanishing magnetic shear are introduced and discussed in Sec. \ref{Theoretical Model} in different parameter regimes, depending on the relative magnitude of $L_{P_E}$ and $\Delta_m$. Detailed numerical investigations and theoretical analysis of the low-frequency SAWs in the presence of EPs are discussed in Sec. \ref{results}, where comparisons between theory and experiments are also made. Finally, conclusions and further discussions are given in Sec. \ref{conclusion}.

\section{The General Fishbone-Like Dispersion Relation for low-frequency SAWs}\label{Theoretical Model}

In this Section, we will present analytical dispersion relations for low-frequency SAW excitation in weakly reversed-shear DIII-D discharges. As stated in the previous Section, two cases determined by the relative magnitude of $L_{P_E}$ and $\Delta_m$ will be used to investigate the low-frequency SAW stability: case I, the local GFLDR model corresponding to $L_{P_E}>\Delta_m $; and case II, the global GFLDR corresponding to $L_{P_E}\simeq\Delta_m $.

Consider case I first. For $L_{P_{E;rel}}>\Delta_m$, the scales of $L_{P_E}$ and $\Delta_m$ can be separated, and the vorticity equation \cite{Chen1994,Tsai1993,Zonca1996,Zonca2014,Zonca2014a,Chen2016} which governs shear Alfv\'en waves (SAWs) can yield the low-frequency electromagnetic fluctuation dispersion relation in the usual local limit, as derived and discussed in great details in Refs. \cite{Tsai1993,Chen1994,Zonca2007,Zonca2014,Zonca2014a,Chen2016}. We just note that, for DIII-D case of interest, the reversed magnetic shear configuration and thermal plasma compression effects should be accounted for properly \cite{Ma2022}. Thus, for $s=0$ at $r_0$ but with finite $S\equiv (r/q)[q^{''}]^{1/2}$, the local GFLDR for low-frequency SAWs can be written as \cite{Zonca2002,Zonca2007,Zonca2014,Zonca2014a,Chen2017}
\begin{equation}\label{zeroshear_GFLDR}
\begin{aligned}
iS(\Lambda_{n}^2-k_{\parallel n0}^2q_{min}^2R_0^2)^{1/2}(1/n)^{1/2}\big[k_{\parallel n 0}q_{min}R_0-i(\Lambda_n^2-k_{\parallel n0}^2q_{min}^2R_0^2)^{1/2}\big]^{1/2}=\delta {\hat W}_{nf}+\delta{\hat W}_{nk}(\omega),
\end{aligned}
\end{equation}
where the generalized inertia term $\Lambda_n(\omega)$ here, including both diamagnetic effects as well as kinetic effects of circulating and trapped particle dynamics, has been derived explicitly in Ref. \cite{Chavdarovski2009} and the main results are summarized in Appendix \ref{A}. The right hand side of Eq. \ref{zeroshear_GFLDR} contains both ``fluid" ($\delta {\hat W}_{nf}$) and ``kinetic" ($\delta {\hat W}_{nk}$) contributions to the potential energy in the ``regular" ideal region. In the low-frequency limits ($|\Lambda_n^2|\ll 1$), $\delta {\hat W}_{nf}$ is independent of the frequency and the explicit expression, specialized to the ($s$, $\alpha$) model equilibrium \cite{Connor1978} with circular flux surfaces, reads,
\begin{equation}\label{dwf}
\delta {\hat W}_{nf}\simeq \frac{\pi}{4}\bigg(\frac{S^2k_{\parallel 0}q_{min}R_0}{n}-\frac{3}{2}\alpha^2 S\big|\frac{k_{\parallel 0}q_{min}R_0}{n}\big|^{1/2}+\frac{9}{32}\alpha^4\bigg)
\end{equation}
where $\alpha=\alpha_c+\alpha_E$, $\alpha_c=-R_0q_{min}^2d\beta/dr$ and $\alpha_E=-\frac{1}{2}R_0q_{min}^2d(\beta_{E \parallel}+\beta_{E\perp})/dr$. Note that Eq. \ref{dwf} includes the contribution of the energetic particle adiabatic and convective responses as well \cite{Chen1984}.

The term $\delta {\hat W}_{nk}$ is always a function of the mode frequency $\omega$, as it reflects resonant as well as non-resonant wave-particle interactions. For simplicity but still relevant to the DIII-D case, we take $F_{0E}$ to be a single pitch angle ($\lambda=\mu/\varepsilon$) slowing-down beam ion equilibrium distribution function; i.e., $F_{0E}=\frac{B_0\beta_{E}(r)}{2^5\sqrt{2}\pi^2m_E\varepsilon_b}\sqrt{(1-\lambda_0B_0)}\varepsilon^{-3/2}\delta (\lambda-\lambda_0)$. Here, $\beta_{E}(r)\equiv 8\pi P_E(r)/B_0^2$ is the ratio of EP kinetic and magnetic pressures and $B_0$ the on-axis equilibrium magnetic field, $\delta(x)$ is the Dirac function, $\mu$ is the magnetic moment and $\varepsilon=\upsilon^2/2 \leq \varepsilon_b$ with $\varepsilon_b$ being the EP birth energy per unit mass. Then the explicit expression of non-adiabatic contribution $\delta {\hat W}_{nku}$ for the passing energetic ions is given by \cite{Tsai1993,Chen1994}
\begin{equation}\label{dwk_u}
\delta {\hat W}_{nku}\simeq  \frac{\pi\alpha_{E}}{2^{5/2}}(1-\lambda_0B_0/2){\bar \omega}\left[2-{\bar \omega}\ln \left( \frac{{\bar \omega}+1}{{\bar \omega}-1}\right)\right],
\end{equation}
where ${\bar \omega}=\omega/\omega_{tEm}$ and $\omega_{tEm}\equiv \sqrt{2\varepsilon_b}/qR_0$ is the EP transit frequency at the maximum particle energy.

It is worthwhile emphasizing that the finite $k_{\parallel n0}q_{min}R_0$ in Eq. \ref{zeroshear_GFLDR} plays an important stabilizing role since it represents the finite line bending effect at $r=r_0$ \cite{Zonca2007,Zonca2014,Zonca2014a}. Furthermore, the expression of $\Lambda_n$ depends on the mode polarization via $S_f\equiv (i\delta E_{\parallel}/k_{\parallel})_{a.c.}\big/\delta \phi_{d.c.}$, where $a.c.$ and $d.c.$ refer to the sinusoidal and nearly constant (flute-like) components of the parallel electric field, wave vector, and scalar potential fluctuation \cite{Chavdarovski2014,Chen2017}. The detailed expression of $S_f$, again, is given in the Appendix \ref{A}. Here, we just note that $|S_f|$ is much smaller than unity for shear Alfv\'en wave and order of unity for ion acoustic wave \cite{Chavdarovski2009,Chavdarovski2014,Chen2017}.

We remark here that, in the moderate pressure gradient case, the local GFLDR for the low-frequency SAWs is enough to delineate the underlying physics of the experimental and simulation results. However, the local GFLDR for the low-frequency SAWs, given by Eq. \ref{zeroshear_GFLDR}, will fail in the presence of strong EP pressure gradient, i.e., case II. In this case, two typical scale lengths $L_{P_{E,cl}}$ and $\Delta_m$ can not be separated anymore and, thus, a global dispersion relation is needed which can be derived from the vorticity equation, i.e., Eq. (1) of Ref. \cite{Zonca2002}. Noting that the mode structure is dominated by single toroidal and poloidal mode numbers, $(n,m)$, the governing equation reads
      \begin{equation}\label{global_dr}
      \begin{aligned}
      ({\bf e}_\theta-{\bf e}_{r}\xi)\cdot \left[\Lambda^2-\Omega^2_{A,m}\left(1+\frac{x^2}{\Omega_{A,m}}+\frac{x^4}{4\Omega^2_{A,m}}\right)\right]({\bf e}_\theta-{\bf e}_{r}\xi)\delta \phi_m-(F+K)\delta\phi_m=0,
      \end{aligned}
      \end{equation}
where ${\bf k}_\perp/k_\theta=-({\bf e_\theta}-{\bf e}_r\xi)$ with ${\bf e}_r$ and ${\bf e}_\theta$ being, respectively, the radial and poloidal unit vectors, $x^2=nq_{min}''(r-r_0)^2$, $\xi\equiv(i/n^{1/2})S(\partial/\partial x)$, and $\delta \phi_m$ is the $m$th poloidal harmonic of the scalar field perturbation. It is worth noting that, toroidal coupling among different poloidal harmonics is typically not important for modes in the reversed magnetic shear region, consistent with the mode being dominated by single $m$ and $n$.
The terms $F$ and $K$ in Eq. \ref{global_dr} represent, respectively, the fluid-like particle and energetic ion contributions with their explicit form reading
\begin{equation}
\begin{aligned}
&F\simeq D_S^2-4\alpha^2D_S+2\alpha D_S^2-(\alpha+1)\alpha+2\alpha^3,\\
&K\simeq\frac{2\pi q_E^2 q^2 R_0^2 \omega}{m_Ec^2}\left\langle\frac{\Omega_{dE}^2QF_{0E}}{\omega_{tE}^2-\omega^2}\right\rangle_\upsilon=\frac{2}{\pi}\delta {\hat W}_{nku},
\end{aligned}
\end{equation}
where $D_S=S\sqrt{\Omega_{A,m}/n}$, $q_E$ and $m_E$ are the electric charge and mass of energetic ions, $\Omega_{dE}=(\upsilon_{E\perp}^2/2+\upsilon_{E\parallel}^2)/\omega_{cE}R_0$,  $\omega_{tE}=\upsilon_{E\parallel}/qR_0$, $QF_{0E}=(\omega\partial_\varepsilon+{\hat \omega}_{\ast E})F_{0E}$, ${\hat \omega}_{\ast E}F_{0E}=\omega^{-1}_{cE}({\bf k} \times {\bf b})\cdot \nabla F_{0E}$, $\omega_{cE}=q_EB/m_Ec$, $\langle(...)\rangle_\upsilon=\int d^3\upsilon (...)$, and the subscripts $\parallel$ and $\perp$ represent the parallel and perpendicular components with respect to the equilibrium magnetic field ${\bf b}$.

\begin{figure}[htbp]
\centering
\includegraphics[width=0.5\textwidth]{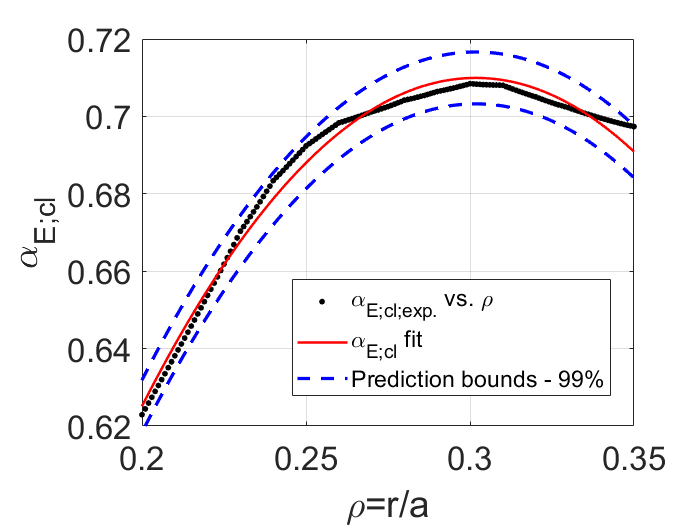}
\caption{The radial dependence of the normalized pressure gradient of EPs with the classical profile. Here, the normalized radial position of $q_{min}$ is $\rho_0\equiv r_0/a=0.28$.}
\label{aE_classical}
\end{figure}

Equation \ref{global_dr} is an ordinary differential equation and, generally, requires a numerical approach to be solved. However, for DIII-D case, the radial dependence of the normalized pressure gradient of energetic ions with the classical profile, as is shown by black curve in Fig. \ref{aE_classical}, can be well fitted by the analytic formula $\alpha_E(\rho)=c_1\left(1-(\rho-c_2)^2/c_3^2\right)$, with $c_1=0.7099$, $c_2=0.3018$ and $c_3=0.2944$. This allows us to obtain simple analytical dispersion relations for low-frequency SAWs excitation. We just note that the maximum drive of energetic ions is located around $\rho=c_2=0.3018$, which deviates from the radial position of $q_{min}$. Then $\alpha_E(r)$ in Eq. \ref{dwk_u} can be rewritten as
      \begin{equation}\label{a_Er}
       \alpha_E(r)=\delta_a\alpha_{E0}\left(1-\frac{(r-r_0+\delta_b)^2}{\delta_c^2L_{PE;cl}^2}\right),
      \end{equation}
where $\delta_a=c_1/\alpha_{E0}$, $\delta_b=r_0-c_2a$  and $\delta_c=c_3a/L_{PE;cl}$, $a$ is the minor radius, $\alpha_{E0}$ and $L_{PE;cl}$ are evaluated at $r=r_0$.
Introducing the notation $x=r-r_0=\sigma z-\delta_b$, Eq. \ref{global_dr} is readily cast into the form
      \begin{equation}\label{Weibel_eq}
      \begin{aligned}
      &\frac{\partial^2}{\partial z^2}\delta \phi_m -\frac{n\sigma^2}{S^2}\left(1-\frac{F+\frac{2\delta_a}{\pi}\delta {\hat W}_{nku0}}{\epsilon_{A0}}\right)\delta \phi_m - \frac{1}{4}z^2\delta \phi_m=0,\\
     & \frac{2n\sigma^4\delta_a \delta {\hat W}_{nku0}}{\epsilon_{A0}\pi S^2\delta_c^2L^2_{PE;cl}}=\frac{1}{4},
      \end{aligned}
      \end{equation}
     where $\epsilon_{A0}=\Lambda^2-\Omega^2_{A,m}$,
         $\delta {\hat W}_{nku0}=\frac{\pi\alpha_{E0}}{4\sqrt{2}}\left[2-{\bar \omega}\ln \left( \frac{{\bar \omega}+1}{{\bar \omega}-1}\right)\right]$.
Then, Eq. \ref{Weibel_eq} yields the following global dispersion relation for low-frequency SAWs,
     \begin{equation}\label{global_L_eq}
     \frac{-n^{1/2}\pi^{1/2}\delta_cL_{PE;cl}\epsilon_{A0}^{1/2}}{2\sqrt{2}S\delta_a^{1/2}\delta{\hat W}_{nku0}^{1/2}}\left(1-\frac{F+\frac{2\delta_a}{\pi}\delta{\hat W}_{nku0}}{\epsilon_{A0}}\right)=2L+1, {\quad\text{$L=0,1,2,3$ ...}}
     \end{equation}
Here, the integer $L$ is the radial eigenmode number.
The corresponding eigenfunction reads
      \begin{equation}\label{dphi_mr}
      \delta\phi_m(r)= H_L(z)e^{-z^2}\propto \exp\left(-\frac{(r-r_0+\delta_b)^2}{4\sigma^2}\right),
      \end{equation}
where $H_L(z)$ represents $L$th order Hermite polynomials and the causality constraints upon the discrete bound modes requiring ${\cal R}e(\sigma^2)>0$, where $\sigma^2$ is solved for from the second of Eqs. \ref{Weibel_eq} consistently with the dispersion relation, Eq. \ref{global_L_eq}. The typical radial width, $w$, of $\delta \phi_m(r)$ is determined by $w^2=4\sigma^2$.

Equations \ref{zeroshear_GFLDR} and \ref{global_L_eq} constitute the results of the present section, i.e., the local and global GFLDR for the low-frequency SAWs excited by energetic ions. With their explicit form, we can compute the individual terms involved in equations and investigate the linear properties of the experimentally observed low-frequency SAWs.

\section{The low-frequency SAW Instabilities Numerical Results and Analysis}\label{results}
In this Section, we separately present numerical results for the local and global low-frequency SAW stability properties in the presence of energetic ions, for which the dispersion relation is given by Eqs. \ref{zeroshear_GFLDR} and \ref{global_L_eq}.
    \begin{figure}[htbp]
    \centering
    \includegraphics[width=0.47\textwidth]{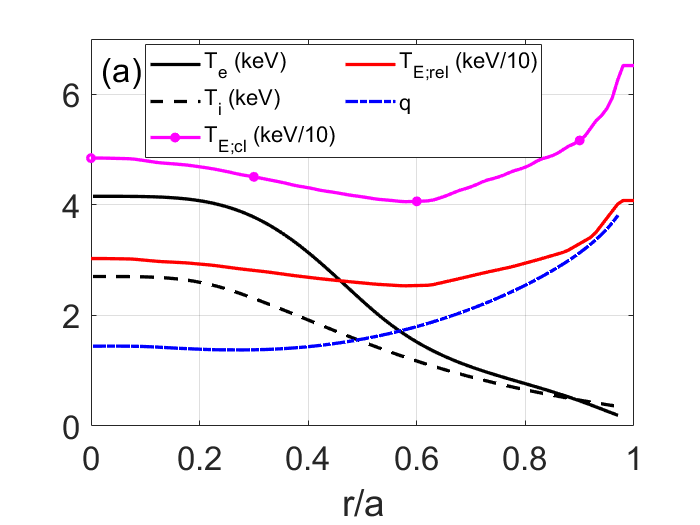}
    \includegraphics[width=0.47\textwidth]{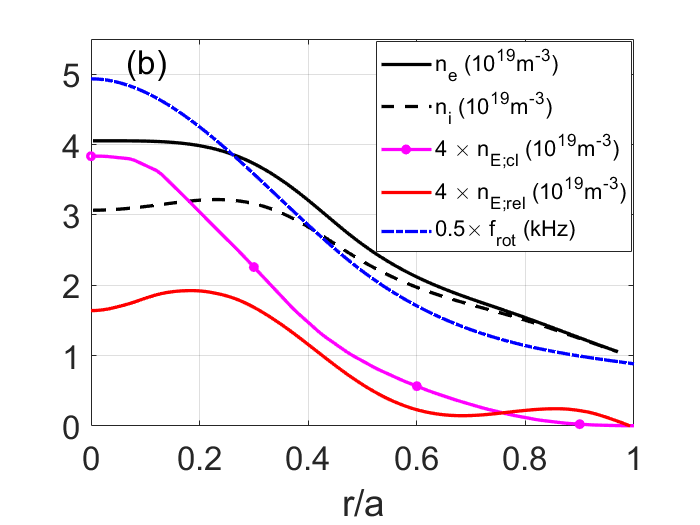}
    \caption{Radial profiles of (a) temperature and $q$ and (b) density and toroidal rotation frequency $f_{rot}$ of DIII-D shot \#178631 used for numerical studies.}
    \label{equili_profiles}
    \end{figure}
The numerical investigations use experimental equilibrium and profiles as shown in Fig. \ref{equili_profiles} for the DIII-D shot \#178631 at the time $t=1200$ ms \cite{Heidbrink2021}, where the $q$-profile has a reversed shear configuration with $q_{min}=1.37$ at $r_0/a=0.28$ and $q_{min}$ decreases from 1.49 to 1.18 in the time window $1050$ ms $<t<1350$ ms, as shown in Fig. 6 (b) in Ref. \onlinecite{Heidbrink2021}.

\subsection{The local low-frequency SAW stability properties}

We first consider the linear properties of the low-frequency SAW with relaxed energetic ion profile, i.e., case I. The local equilibrium parameters used in the numerical studies evaluated at $r_0/a=0.28$ are $S=0.5895$, $\tau=T_e/T_i=$3.86 keV/2.37 keV=1.62, $n_{e}=3.80\times10^{19}$ $m^{-3}$, $n_{i}=3.19\times10^{19}$ $m^{-3}$, $\epsilon_r=r_0/R=0.10$, $\beta_i\simeq0.01$, $\epsilon_{ni}=L_{ni}/R_0=0.414$,
$\eta_i=L_{ni}/L_{Ti}=0.8324$, $\omega_{\ast ni}/\omega_{ti}=0.1919$, $(m,n)=(8,6)$, $k_\theta \rho_{Li}=0.2555$ and $k_\theta \rho_{Le}=0.0054$. Other fixed equilibrium parameters are $a=0.64$ m, $R_0=1.74$ m, $B_0=1.8$ T. Here, $k_\theta$ is the poloidal wavenumber, $\rho_{Li}$ and $\rho_{Le}$ are the Larmor radii of thermal ions and thermal electrons, respectively.
\begin{figure}[htbp]
\centering
\includegraphics[width=0.47\textwidth]{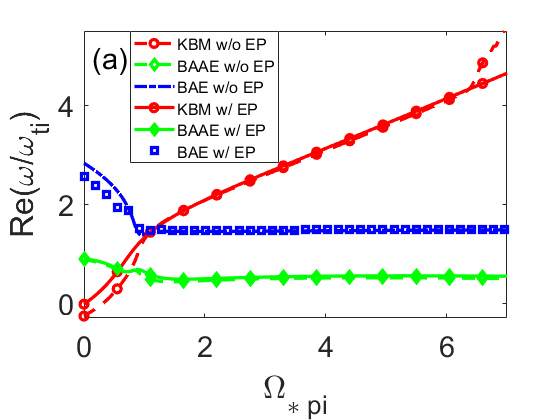}
\includegraphics[width=0.47\textwidth]{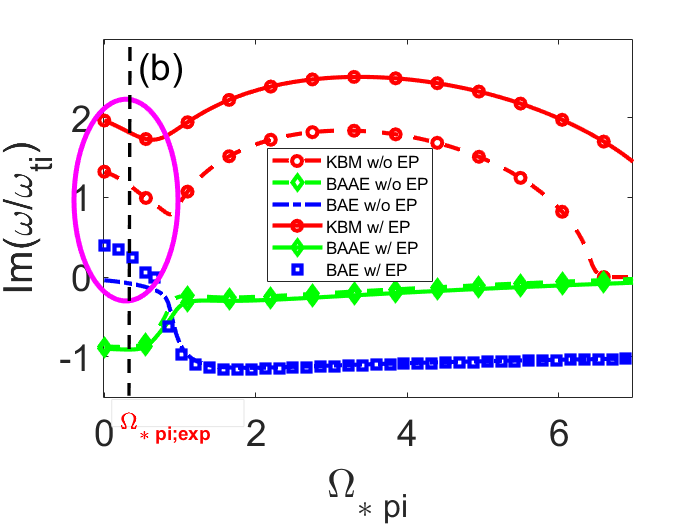}
\includegraphics[width=0.47\textwidth]{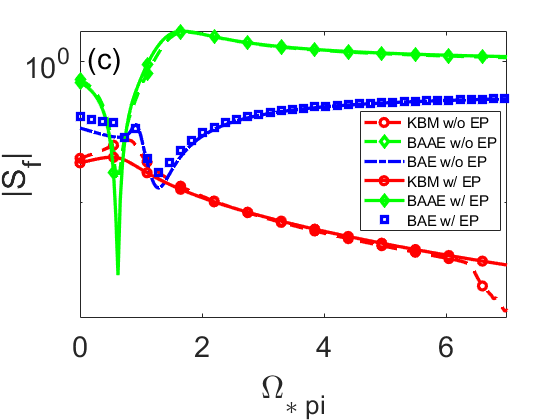}
\caption{Dependence of the (a) real frequencies, (b) growth rates and (c) polarization of the low-frequency SAWs on $\Omega_{\ast pi}\equiv \omega_{\ast pi}/\omega_{ti}$ for the cases without (w/o) and with (w/) EP effects. Here, a dashed vertical line represents the experimental value of $\Omega_{\ast pi;exp}$ of about 0.35.}
\label{changewpi}
\end{figure}

\begin{figure}[htbp]
\includegraphics[width=0.46\textwidth]{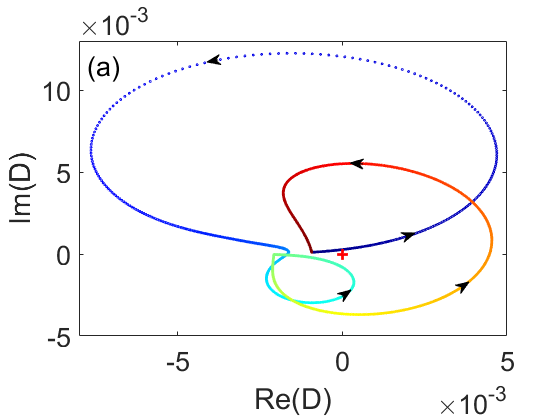}
\includegraphics[width=0.46\textwidth]{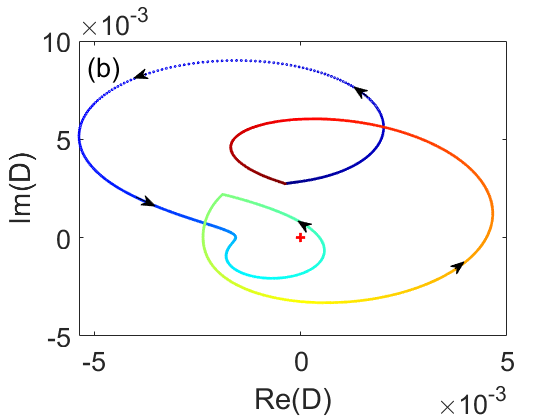}
\caption{{{The Nyquist diagram in the complex ${\rm D}(\omega)$ plane for the cases (a) without and (b) with EP effects.}}}
\label{nyquist}
\end{figure}

Dependencies of the (a) mode frequencies, (b) growth rates and (c) mode polarization predicted by Eq. \ref{zeroshear_GFLDR} are shown in Fig. \ref{changewpi} as a function of the normalized thermal ion diamagnetic frequency $\Omega_{\ast pi}\equiv \omega_{\ast pi}/\omega_{ti}$ for the cases without and with the consideration of EP effects. According to the scaling of mode frequencies with physical parameters and the value of the $|S_f|$ \cite{Chavdarovski2014}, three branches in Fig. \ref{changewpi} can be classified as: (i) the KBM (red curves marked with circles), with a frequency scaling with $\omega\sim \omega_{\ast pi}$; (ii) the BAE (blue curves), with the frequency being close to the well-known estimate $\omega/\omega_{ti}=q_{min}\sqrt{7/4+\tau}\simeq 2.51$; and (iii) the BAAE (green curves marked with diamonds), with a frequency of about half of the BAE and experiencing strong damping. The EP effects on the low-frequency SAW stabilities are apparent in the region highlighted by the purple curve of Fig. \ref{changewpi} (b), where the KBM is the only unstable mode in the absence of EPs, while both the KBM and BAE are unstable in the low-frequency region in the presence of EPs. In particular, the diamagnetic ion frequency calculated on the basis of experimental parameters is $\Omega_{\ast pi;exp}=0.3517$, as shown by the dashed vertical line. In this case, both KBM and BAE are unstable with the frequencies in the plasma frame being 5.6 kHz and 63.7 kHz, respectively, which are in good agreement with the experimental observations. Meanwhile, the polarization plot of Fig. \ref{changewpi} (c) shows that KBM and BAE have small values for $|S_f|\lesssim 0.1$, which indicates that the KBM and BAE are essentially of Alfv\'enic polarization. Moreover, in order to exclude the spurious nonzero solutions produced by singularities of the transcendental function of the local GFLDR (${\rm D}$), the Nyquist diagram in the complex D plane presented in Fig. \ref{nyquist} shows that in the presence of EPs, the path encircles the origin twice (see Fig. \ref{nyquist} (b)) but only once without EPs (see Fig. \ref{nyquist} (a)), thus confirming there are two unstable modes with EPs.
 It should be noted that, compared with the frequency insensitive to the EP effects, the growth rate of the KBMs changes significantly in the cases with and without EP effects. This occurs because in our theoretical model the adiabatic and convective contribution of EPs modifies the value of $\delta {\hat W}_f$ via $\alpha$, as is shown in Eq. \ref{dwf}. At this point, in order to obtain more convincing comparison of theoretical prediction and experimental observation, it is necessary to provide a more precise theoretical model and also a more comprehensive experimental analysis.
We also note here that, in this case, the stability/property of the BAAE is not affected by energetic ions --- as is shown by the green dashed lines with symbols (without EP effects) and solid lines with symbols (with EP effects) which are apparently overlaying in all three graphs --- even though it becomes weakly damped by coupling with the KBM due to diamagnetic and trapped particle effects for sufficiently strong $\Omega_{\ast pi}$. The numerical results are consistent with the numerical simulation results reported in Refs. \cite{Lauber2013a,Bierwage2017,Choi2021a} and the theoretical prediction in Ref. \cite{Chen2017}, that is, ``{\it EPs preferentially excite the BAE over the BAAE branch due to the stronger wave-EP interaction}".

\begin{figure}[htbp]
\centering
\includegraphics[width=1.0\textwidth]{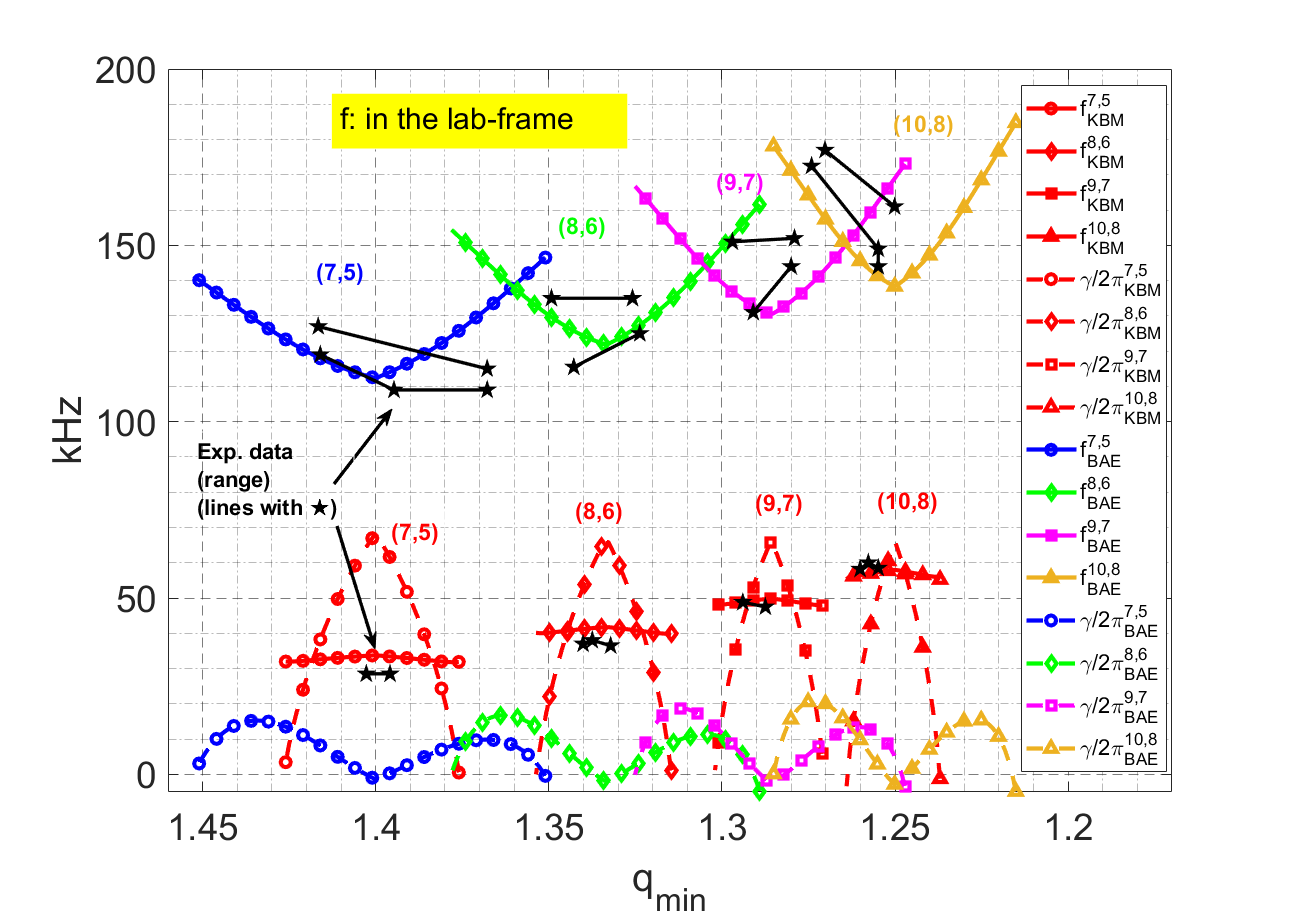}
\caption{Dependence of mode frequencies (solid curves with markers) and growth rates (dashed curves with markers) on $q_{min}$ of the KBMs (red curves) and the BAEs (blue, green, purple and orange curves) for different ($m$, $n$). The experimentally observed frequencies are also shown. For the BAE, since the modes span a range of frequencies, the lines indicate the upper and lower limits of the unstable bands; for the LFAM, the experimental frequency variation is $<0.5$~kHz. In the abscissa, the experimentally measured $q_{min}(t)$ fit  shown in Fig.~8 of \cite{Heidbrink2021a} is used to convert time to $q_{min}$, with an associated uncertainty of $\Delta q_{min}\simeq0.01$. In the ordinate, the theoretical lab-frame frequency incorporates a Doppler shift to the calculated plasma-frame frequency of $nf_{rot}$, with an associated uncertainty of $\sim0.5\times n$~kHz.}
\label{change_qmin}
\end{figure}

We now investigate the underlying instability mechanisms of the ascending spectrum of the higher frequency BAEs and LFAMs observed in DIII-D (see Fig. 8 of Ref. \cite{Heidbrink2021a}) by using $q_{min}$ as the scanning parameter. Figure \ref{change_qmin} shows the dependence of the mode frequencies (solid curves with markers) and growth rates (dashed curves with markers) on $q_{min}$ of the KBMs (red curves) and the BAEs (blue, green, purple and orange curves) for different poloidal and toroidal mode numbers ($m$, $n$).
It is shown that the modes in ascending pattern of higher frequency BAEs and lower frequency KBMs are both separated by approximately $f_{rot}$ of about 7.5 kHz. More specifically, for KBMs, the instabilities peak exactly at the rational values of $q_{min}$; while the BAEs occur at times near rational values of $q_{min}$ but the timing of unstable modes is less precise than for KBMs. In addition, the low-$n$ BAEs deviate more from rational $q_{min}$ crossings than higher $n$ modes. The comparison of the theoretically predicted frequencies with the experimentally measured values can also be seen clearly from Fig. \ref{change_qmin}. As discussed in more detail in the next section, these numerical results are in good agreement with the experimental observations.

In order to gain insight into the different excitation mechanisms of the instabilities presented in Fig. \ref{change_qmin}, let us further analyze the GFLDR in the high-frequency ($|\omega| \gg \omega_{ti}$) and low-frequency $|\omega| \ll \omega_{bi}$ limits.

For $|\omega|\gg |\omega_{ti}|$, the corresponding inertia term of the BAE can be reduced to the simplified expression with $\Lambda^2\simeq\frac{\omega^2-\omega^2_{BAE}}{\omega^2_A}$ \cite{Zonca1996,Zonca2007,Zonca2009}. Here, $\omega^2_{BAE}=(7/4+\tau)\upsilon_i^2/R_0^2$ is the fluid limit expression of the BAE frequency. Taking $\omega=\omega_r+i\gamma$ and $\delta {\hat W}_{ku}={\rm Re}\delta{\hat W}_{ku}+i{\rm Im}\delta{\hat W}_{ku}$, and assuming $|\gamma/\omega_r|$, we have $|{\rm Im}\delta{\hat W}_{ku}/{\rm Re}\delta{\hat W}_{ku}|\ll 1$. Then, for the gap mode, the existence condition is $\delta {\hat W}_{nf}+{\rm Re}(\delta {\hat W}_{nk}(\omega_r))<0$ and the real mode frequency is given by
\begin{equation}\label{BAE_freq}
\omega_r^2=\omega_{BAE}^2\left[1+\frac{\omega_A^2}{\omega_{BAE}^2}\left(k_{\parallel n0}^2q_{min}^2R_0^2-\frac{n}{\left|k_{\parallel n0}q_{min}R_0\right|}\frac{\left(\delta {\hat W}_{nf}+{\rm Re}(\delta {\hat W}_{nk}(\omega_r))\right)^2}{S^2}\right)\right],
\end{equation}
while the growth rate is obtained from
\begin{equation}\label{BAE_growth}
\gamma=-{\rm Im}(\delta {\hat W}_{nk}(\omega_r))\frac{\omega_A^2}{\omega_r}\frac{n\left(\delta {\hat W}_{nf}+{\rm Re}({\delta {\hat W}_{nk}(\omega_r))}\right)}{\left|k_{\parallel n0}q_{min}R_0\right|S^2},
\end{equation}
It can be readily obtained from Eq. \ref{BAE_freq} that the BAE frequency is positively correlated with $\left|k_{\parallel n0}q_{min}R_0\right|$. Therefore, the more deviation from the rational $q_{min}$ surface is, the larger the BAE frequency is, as is shown in Fig. \ref{change_qmin}. Note also that the BAE has a positive frequency. Equation \ref{BAE_growth} imposes ${\rm Im}(\delta {\hat W}_{nk}(\omega_r))>0$ for BAE excitation by EPs via resonant wave-particle interaction.
It can be concluded that the duration of BAEs is influenced by the associated resonances with the EPs, as well as by the value of $q_{min}$ \cite{Heidbrink2021a}.

Similarly, for KBM with $|\omega|\ll |\omega_{bi}|$, we have $\Lambda^2\simeq c_0\frac{q_{min}^2}{\sqrt{\epsilon}}\frac{(\omega-\bar{\omega}_{di})(\omega-\omega_{\ast pi})}{\omega^2_A}$ \cite{Zonca2007,Chavdarovski2009,Chavdarovski2014,Chen2020,Heidbrink2021}. Here, $\bar{\omega}_{di}$ is the average thermal-ion precession frequency, $c_0\simeq 1.6$ due to trapped and barely circulating particles \cite{Rosenbluth1998,Graves2000}. Thus, the real mode frequency is given by
\begin{equation}
\omega= \frac{1}{2}({\bar \omega}_{di}+\omega_{\ast pi})\pm \frac{1}{2}\left[(\omega_{\ast pi}-{\bar \omega}_{di})^2-\frac{4\omega_A^2\sqrt{\epsilon}}{q_{min}^2c_0}\left(\frac{n\left(\delta {\hat W}_{nf}+{\rm Re}({\delta {\hat W}_{nk}(\omega_r))}\right)^2}{\left|k_{\parallel n0}q_{min}R_0\right|S^2}-k_{\parallel n0}^2q_{min}^2R_0^2\right)\right]^{1/2},
\end{equation}
and the system is reactively unstable if
   \begin{equation}\label{reactive_condition}
   \frac{|\omega_{\ast pi}-{\bar \omega}_{di}|^2}{\omega_A^2}<\frac{4\sqrt{\epsilon}}{q_{min}^2c_0}\left(\frac{n\left(\delta {\hat W}_{nf}+{\rm Re}({\delta {\hat W}_{nk}(\omega_r))}\right)^2}{\left|k_{\parallel n0}q_{min}R_0\right|S^2}-k_{\parallel n0}^2q_{min}^2R_0^2\right).
   \end{equation}
Note that $\delta {\hat W}_{f}+{\rm Re}\delta {\hat W}_{ku}<0$, due to, again, the causality constraint. Therefore, for the reactive-type instability, the maximum drive sets in when $k_{\parallel n0}q_{min}R_0\rightarrow 0$, which corresponds to the unstable KBM exactly peaking at the rational values of $q_{min}$.

The above numerical results and theoretical analyses have explained the experimental observations that the BAEs deviate more from the rational $q_{min}$ values temporally, compared with the KBM. To further delineate this deviation and its impact on the radial mode structure, numerical investigation of the global model for low-frequency SAWs is needed.

\subsection{The global low-frequency SAW stability properties}
In this part, we consider the case II and apply Eq. \ref{global_L_eq} to investigate the global low-frequency SAW stability properties with the classical energetic ion profile.
\begin{figure}[htbp]
\centering
\includegraphics[width=0.47\textwidth]{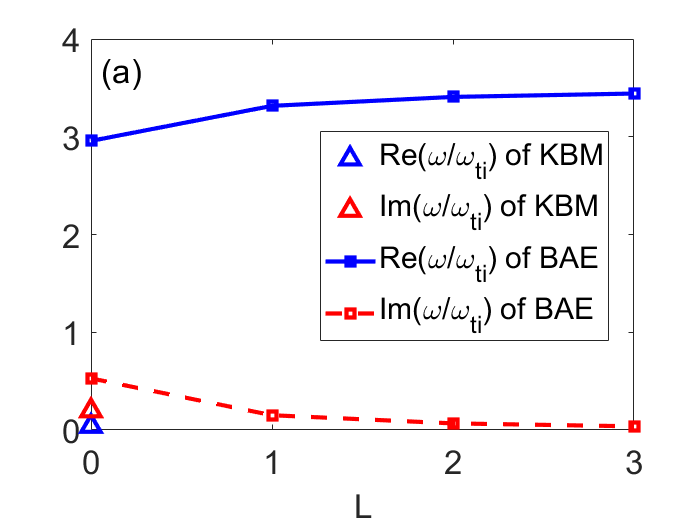}
\includegraphics[width=0.47\textwidth]{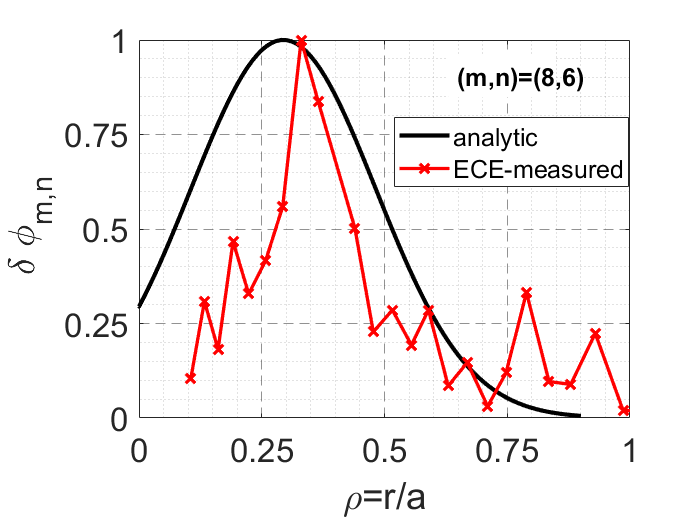}
\caption{(a) Dependence of the real frequencies (blue markers) and growth rates (red markers) of the KBM (triangle markers) and BAE (line with markers) on the radial mode number $L$; (b) the radial mode structure $\delta \phi_{m}$(r) for the $L=0$ BAE. The approximate experimental measurement of the mode structure of BAE is also shown.}
\label{change-L}
\end{figure}

Figure \ref{change-L} shows (a) the dependence of the real frequencies (blue markers) and growth rates (red markers) of the KBM (triangle markers) and BAE (line with markers) on the radial mode number $L$; and (b) the radial mode structure $\delta \phi_{m}$(r) for the $L=0$ BAE. It can be found that (i) the ground eigenstate with $L=0$ is most unstable for the BAE and KBM; (ii) for BAE, the frequency and growth rate in the plasma frame is $(80.7+15.2i)$ kHz with the ratio of the growth rate to real frequency $\gamma/\omega\simeq 0.19$, which is the typical feature of the marginally unstable gap mode excited by EPs; and (iii) for KBM, the frequency and growth rate in the plasma frame is $(-3.2+5.7i)$ kHz with $\gamma/\omega\simeq 1.8$, which is the typical feature of the reactive-type instability, consistent with the results reported in Ref. \cite{Choi2021a}.

Correspondingly, the radial eigenfunction plot of the BAE for $L=0$, as shown in Fig. \ref{change-L} (b), presents that $\delta \phi_m$ has a Gaussian form with a shape similar to the experimentally measured radial mode structure. In this case, the radial width of $\delta \phi_m$ by theory is $w= 0.2107$, is comparable to the scale length of energetic-ion pressure, i.e., $L_{P_{E;cl}}=0.1773$; consistent with the analysis of Fig. \ref{LP_scales}. Note that determined by the EP distribution, the BAE eigenfunction peaks at the radial position of the maximum energetic particle pressure gradient, resulting in a large deviation from the $q_{min}$ surface. It can also be expected that the KBM eigenfunction should peak at the rational values of $q_{min}$ where the instability drive is maximum.

\begin{figure}[htbp]
\centering
\includegraphics[width=0.495\textwidth]{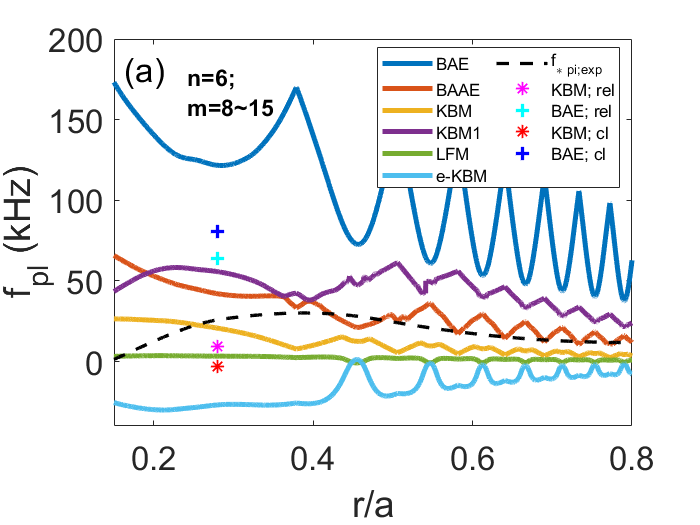}
\includegraphics[width=0.495\textwidth]{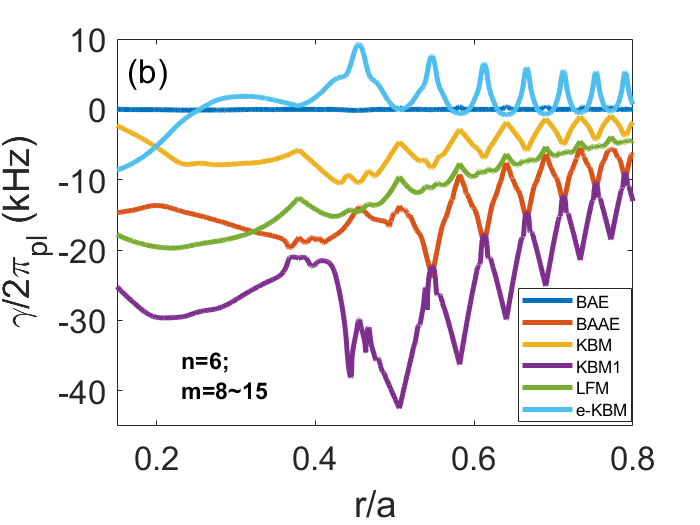}
\includegraphics[width=0.495\textwidth]{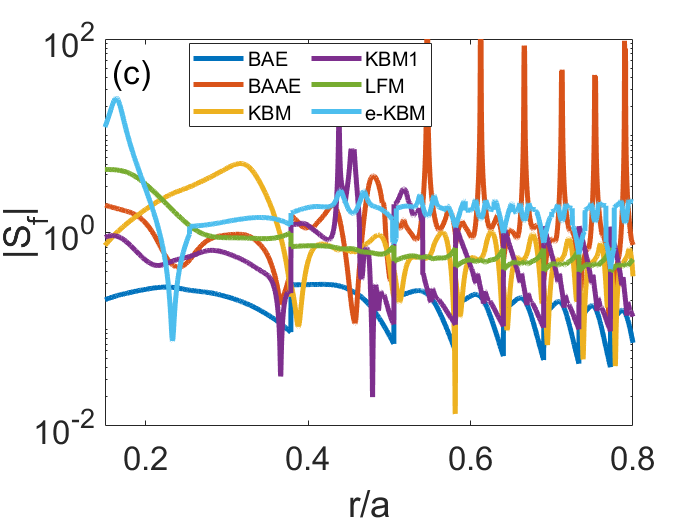}
\caption{The continuous spectra of low-frequency shear Alfv\'en and acoustic branches for n=6, m=8-15. The equilibrium profiles of DIII-D \#178631 at 1200 ms are adopted.}
\label{n6_continuum_spectra}
\end{figure}

Finally, the continuous spectra plots for low-frequency shear Alfv\'en and acoustic waves given by $\Lambda_n^2(\omega)=k_{\parallel n}^2 q^2R_0^2=(nq-m)^2$ \cite{Zonca1996,Zonca2009,Zonca2010,Zonca2014,Zonca2014a,Falessi2019a,Falessi2020} are shown in Fig. \ref{n6_continuum_spectra}. Here, the inertia term includes the diamagnetic effects and thermal ion compressibility as well as drift Alfv\'en wave and drift wave sideband coupling via the wave-thermal-passing-ion interaction and diamagnetic effect \cite{Zonca2010}. The figure shows that based on the GFLDR, the nature of various branches can be clearly classified via their frequencies (a), growth rates (b) and  polarizations (c). Here, the short notation ``e-KBM" represents the branch of the KBM propagating in the thermal-electron diamagnetic drift direction. The unstable continuum spectrum of the e-KBM is due to the inclusion of the kinetic dynamics of thermal particles in inertia term. In addition, the frequencies of the $(m,n)=(8,6)$ BAE and the $(m,n)=(8,6)$ KBM calculated by the local and global cases are, respectively, in the gaps of the BAE and KBM continua, which is consistent with the numerical simulation results reported in Refs. \cite{Heidbrink2021,Choi2021a}.

\section{Summary and Discussions}\label{conclusion}
The present work has addressed linear properties of the low-frequency shear Alfv\'en waves (SAWs) with the consideration of energetic ions in DIII-D reversed magnetic shear tokamak experiments. By analyzing the experimental equilibrium profiles, the local and global models for low-frequency SAWs for weak and/or vanishing magnetic shear are discussed based on the unified theoretical framework of the generalize fishbone-like dispersion relation (GFLDR). Resorting to numerical and theoretical analyses, the dependences of mode frequency, growth rate and polarization on the minimum of the safety factor ($q_{min}$), as well as the instability mechanisms are delineated.

The main results of this work are that the LFAMs and BAEs observed in DIII-D experiments are, respectively, the reactive-type and dissipative-type unstable modes with predominantly Alfv\'enic polarization. Due to the different instability mechanisms, BAE peak occurs further away from the rational $q_{min}$ than LFAM peak does. The BAE eigenfunction is localized at the radial position with the strongest energetic-ion-drive spatially, which leads to deviation from the radial position of $q_{min}$.

The theoretical analysis explains many experimental observations.
\begin{enumerate}
\item The theory successfully explains the temporal pattern of two bands of instability, the BAE band and the LFAM band, that both appear near rational values of $q_{min}$ but with distinctly different stability properties.
\item The predicted values of KBM frequency are in excellent agreement with the experimental LFAM frequencies. The KBM can be unstable even in the absence of energetic particles (EPs).
\item The predicted values of BAE frequency span the same range as the experimentally observed values.
\item The theory also successfully explains the absence of a third branch of instability at BAAE frequencies, as that branch is predicted to be stable.
\item Experimentally, an individual unstable BAE spans a much larger range of frequencies than an unstable LFAM, another feature successfully reproduced by theory.
\item Experimentally, unstable LFAMs only persist for a few milliseconds. The short duration of the LFAM is consistent with the very strong $q_{min}$ dependence of the KBM growth rate.
\item In experiment, unstable BAEs persist longer than LFAMs, which is consistent with the weaker dependence of the BAE growth rate on $q_{min}$ in theory.
\item Temporally, in experiment, LFAMs occur at rational values of $q_{min}$; BAEs also occur near rational values but less precisely. This feature is also reproduced by the theoretical stability predictions: the KBM growth rate peaks sharply at rational $q_{min}$ values but the peak of the BAE growth rate deviates slightly.
\item In experiment, for both the LFAM and the BAE, unstable modes with higher values of toroidal mode number $n$ are of shorter duration than lower values of $n$. The narrower growth rate curves as $n$ increases successfully explains this feature.
\item Experimentally, the BAE radial eigenfunction has an approximately gaussian shape, consistent with the theoretical prediction that the $L=0$ radial harmonic is most unstable.
\item Experimentally, the LFAM is more unstable in plasmas with hydrogen than in pure deuterium plasmas \cite{Heidbrink2021b}, a feature explained by the higher value of $\omega_A$ in hydrogen plasmas. As Eq. \ref{reactive_condition} shows, a larger value of $\omega_A$ lowers the instability threshold.
\end{enumerate}

On the other hand, there are three discrepancies between theory and experiment.
\begin{enumerate}
\item Although the predicted KBM growth rate correctly peaks sharply for rational values of $q_{min}$, it remains positive for a much longer duration than the LFAMs are observed experimentally. Evidently, an additional damping mechanism is missing in the theory.
\item Although the predicted KBM growth rate has changed significantly for the cases with and without EPs, there is no apparent dependence of LFAM stability on EPs experimentally. Therefore, a more precise theoretical model and more comprehensive experimental analysis are needed for meaningful comparison.
\item Although the predicted BAE frequency spans the observed values, the predicted frequency has a parabolic shape with time, while the experimental frequency has a less regular shape. A likely explanation for this discrepancy is imprecise modeling of the fast-ion distribution function.
\end{enumerate}

Finally, there is one theoretical prediction that is inconclusive experimentally: the mode polarization. Theory predicts predominately Alfv\'enic polarization for both the KBM and the BAE. In experiment, low toroidal mode number ($n\leq 3$) BAEs are usually observed on external magnetic coils; LFAMs are never detected, but the inferred toroidal mode numbers typically span a larger range than those normally detected for RSAEs or BAEs. DIII-D is equipped with one diagnostic that can detect internal magnetic fields, a radial interferometer-polarimeter (RIP) \cite{Chen2016b} that measures the line integral of the density and radial magnetic field, $\int n_e B_r dl$. This diagnostic clearly detects RSAEs and BAEs, which is consistent with their expected shear-wave polarization. Fluctuations are observed by RIP for some LFAMs, indicating that there is at least some magnetic component, but the signal is weaker than for RSAEs and BAEs. It is not presently known if this difference is due to a line-integral effect associated with the mode structure or if the LFAM polarization is less Alfv\'enic than the other modes.

\section*{Acknowledgments}
One of authors (R.R. Ma) would like to acknowledge Dr. Lei Yang and Dr. Yunpeng Zou for their useful discussions and the DIII-D team for providing the experimental data. The authors thank Dr. Xiaodi Du for helpful comments concerning the mode polarization.
R.R Ma is also grateful to the \href{https://www.afs.enea.it/zonca/CNPS/}{{\color{black}\it{Center for Nonlinear Plasma Science }}} (CNPS) for its enlightening academic discussion, which provides a valuable sources of scientific stimuli.

This work has been supported in part by the National key R\&D Program of China under Grant Nos. 2022YFE03040002 and 2018YFE0304103, by the National Science Foundation of China under Grant Nos. 12261131622 and 12175053 and Natural Science Foundation of Sichuan under Grant No. 2022NSFSC1814 and Sichuan Science and Technology Program under Grant No. 2022ZYD0019. This work has also been carried out within the framework of the EUROfusion Consortium, funded by the European Union via the Euratom Research and Training Programme (Grant Agreement No. 101052200 -- EUROfusion). Views and opinions expressed are however those of the author(s) only and do not necessarily reflect those of the European Union or the European Commission. Neither the European Union nor the European Commission can be held responsible for them.
This material is based upon work supported by the U.S. Department of Energy, Office of Science, Office of Fusion Energy Sciences, using the DIII-D National Fusion Facility, a DOE Office of Science user facility, under Awards DE-FC02-04ER54698 and DE-SC0020337.

{\small This report was prepared as an account of work sponsored by an agency of the United States Government. Neither the United States Government nor any agency thereof, nor any of their
employees, makes any warranty, express or implied, or assumes any legal liability or responsibility for the accuracy, completeness, or usefulness of any information, apparatus, product, or process disclosed, or represents that its use would not infringe privately owned rights. Reference herein to any specific commercial product, process, or service by trade name, trademark, manufacturer, or otherwise, does not necessarily constitute or imply its endorsement, recommendation, or favoring by the United States Government or any agency thereof. The views and opinions of authors expressed herein do not necessarily state or reflect those of the United States Government or any agency thereof.}

\appendix

\section{Detailed Expressions of $\Lambda_n^2$ and $S_f$}\label{A}
Detailed derivations of the generalized inertia, $\Lambda_n^2$ and wave polarization, $S_f$, can be found in Ref. \onlinecite{Chavdarovski2009}. Here, we only present the results.
In low-$\beta$ ($\beta=8\pi P/B_0^2\approx \epsilon^2$) axisymmetric tokamak plasmas,
\begin{equation}\label{lambda_2009}
\begin{aligned}
\Lambda_n^2=I_{\phi}\left[\frac{\omega^2}{\omega_A^2}\big(1-\frac{\omega_{\ast pi}}{\omega}\big)
+\Lambda^2_{cir}+ \Lambda^2_{tra}\right],
\end{aligned}
\end{equation}
where $\Lambda^2_{cir}$ and $\Lambda^2_{tra}$ represent, respectively, the modified circulating and trapped ion responses, and $I_\phi$ describes the non-vanishing `flute-like' component of the parallel electric field ($\delta E_\parallel$) due to the effect of trapped thermal particle precession resonance  \cite{Chavdarovski2009,Chavdarovski2014}. Meanwhile, $\omega_A=\upsilon_A/qR_0$ is the Alfv\'en frequency with $\upsilon_A$ being the Alfv\'en velocity, and $\omega_{\ast ps}=(T_sc/e_sB)({\textbf{\emph k}}\times{\textbf{\emph b}})\cdot ({\bf \nabla}n_s/n_s+{\bf \nabla} T_s/T_s)\equiv \omega_{\ast ns}+\omega_{\ast Ts}$ is the thermal particle diamagnetic drift frequency due to density and temperature gradients.

For $\Lambda_n^2$, the various terms involved in Eq. \ref{lambda_2009} are given by  \cite{Chavdarovski2009}
\begin{equation}\label{Lambda_cir}
\begin{aligned}
\Lambda^2_{cir}=q^2\frac{\omega\omega_{ti}}{\omega^2_A}\Big[\Big( 1-
\frac{\omega_{\ast ni}}{\omega}\Big)\Big( & F\Big(\frac{\omega}{\omega_{ti}}\Big) + \Delta F\Big(\frac{\omega}{\omega_{ti}} \Big)\Big)-\frac{\omega_{\ast Ti}}{\omega}\Big( G\Big(\frac{\omega}{\omega_{ti}}\Big)+ \Delta G \Big( \frac{\omega}{\omega_{ti}}\Big)\Big)\\
&+\frac{\omega\omega_{ti}}{4{\bar \omega}_{Di}^2}\Big( N_1\big(\frac{\omega}{\omega_{ti}}\Big) + \Delta N_1\Big( \frac{\omega}{\omega_{ti}}\Big)\Big){S_f}(\omega,{\bar\omega}_{Di},\omega_{bi},\omega_{ti})
\Big],
\end{aligned}
\end{equation}
\begin{equation}\label{Lambda_tra}
\Lambda^2_{tra}=\frac{\omega^2\omega_{bi}^2}{\omega_A^2{\bar \omega}_{Di}^2}\frac{q^2}{\sqrt{2\epsilon}}\big[P_3+(P_2-P_3){S_f}(\omega,{\bar \omega}_{Di},\omega_{bi},\omega_{ti})\big],
\end{equation}
\begin{equation}\label{I_phi}
I_\phi=1+\frac{\sqrt{2\epsilon}(L(\omega/{\bar \omega}_{Di})+\tau^{-1}L(\omega/{\bar \omega}_{De}))}{1+\tau\omega_{\ast ni}/\omega+\sqrt{2\epsilon}\tau[1-\omega_{\ast ni}/\omega-M(\omega/{\bar \omega}_{Di})-\tau^{-1}M(\omega/{\bar \omega}_{De})]},
\end{equation}
and, as to $S_f\equiv (i\delta E_{\parallel}/k_{\parallel})_{a.c.}\big/\delta \phi_{d.c.}$, it is given by \cite{Chavdarovski2009}
\begin{equation}\label{S_factor}
{S_f}=-\frac{N_1\left(
\frac{\omega}{\omega_{ti}}\right)+\Delta N_1\left(\frac{\omega}{\omega_{ti}}\right)+\sqrt{2\epsilon}P_2}{1+\frac{1}{\tau}+D_1\left(\frac{\omega}{\omega_{ti}}\right)+\Delta D_1\left(\frac{\omega}{\omega_{ti}}\right)+\sqrt{2\epsilon}\left(P_1-P_2\right)}
\end{equation}
where the functions $F(x)$, $\Delta F(x)$, $G(x)$, $\Delta G(x)$, $N_1(x)$, $\Delta N_1(x)$, $D_1(x)$, $\Delta D_1(x)$, $P_1$, $P_2$, $P_3$, $L(\omega/{\bar\omega}_{Ds})$ and $M(\omega/{\bar\omega}_{Ds})$ with $x=\omega/\omega_{ti}$, and using the plasma dispersion function $Z(x)$, are defined as
\begin{equation}
\begin{aligned}
&Z(x)=\pi^{-1/2}\int _{-\infty}^{\infty}\frac{e ^{-y^2}}{y-x}dy,\\
&F(x)=x(x^2+3/2)+(x^4+x^2+1/2)Z(x),\\
&\Delta F(x)=\frac{1}{\pi^{1/2}}\int_0^{\infty}e^{-y}\ln{\left(\frac{x+\sqrt{2\epsilon y}}{x-\sqrt{2\epsilon y}}\right)}\frac{y^2}{4}dy,\\
&G(x)=x(x^4+x^2+2)+(x^6+x^4/2+x^2+3/4)Z(x),\\
&\Delta G(x)=\frac{1}{\pi^{1/2}}\int_0^{\infty}e^{-y}\ln{\left(\frac{x+\sqrt{2\epsilon y}}{x-\sqrt{2\epsilon y}}\right)}\frac{y^2}{4}\left(y-\frac{3}{2}\right)dy,\\
&N_1(x)=2\frac{{\bar\omega}_{Di}}{\omega_{ti}}\left\{\left(1-\frac{\omega_{\ast ni}}{\omega}\right)[x+(1/2+x^2)Z(x)]-\frac{\omega_{\ast Ti}}{\omega}[x(1/2+x^2)+(1/4+x^4)Z(x)]\right\},\\
&\Delta N_1(x)=\frac{{\bar\omega}_{Di}/\omega_{ti}}{\pi^{1/2}}\int_0^{\infty}ye^{-y}\ln{\left(\frac{x+\sqrt{2\epsilon y}}{x-\sqrt{2\epsilon y}}\right)}\left[1-\frac{\omega_{\ast ni}}{\omega}-\frac{\omega_{\ast Ti}}{\omega}\left(y-\frac{3}{2}\right)\right]dy,\\
&D_1(x)=x\left(1-\frac{\omega_{\ast ni}}{\omega}\right)Z(x)-\frac{\omega_{\ast Ti}}{\omega}[x+(x^2-1/2)Z(x)],\\
&\Delta D_1(x)=\frac{{\bar\omega}_{Di}/\omega_{ti}}{\pi^{1/2}}\int_0^{\infty}e^{-y}\ln{\left(\frac{x+\sqrt{2\epsilon y}}{x-\sqrt{2\epsilon y}}\right)}\left[1-\frac{\omega_{\ast ni}}{\omega}-\frac{\omega_{\ast Ti}}{\omega}\left(y-\frac{3}{2}\right)\right]dy,\\
&P_1=-2\frac{\omega^2}{{\bar\omega}_{Di}^2}\left[\left(1-\frac{\omega_{\ast ni}}{\omega}+\frac{3}{2}\frac{\omega_{\ast Ti}}{\omega}\right)G_2-\frac{\omega_{\ast Ti}}{\omega}G_4\right],\\
&P_2=-2\frac{\omega}{{\bar\omega}_{Di}}\left[\left(1-\frac{\omega_{\ast ni}}{\omega}+\frac{3}{2}\frac{\omega_{\ast Ti}}{\omega}\right)G_4-\frac{\omega_{\ast Ti}}{\omega}G_6\right],\\
&P_3=-2\left[\left(1-\frac{\omega_{\ast ni}}{\omega}+\frac{3}{2}\frac{\omega_{\ast Ti}}{\omega}\right)G_6-\frac{\omega_{\ast Ti}}{\omega}G_8\right],\\
&G_n=\frac{1}{\pi^{1/2}}\int_{-\infty}^{\infty}\frac{e^{-x^2}x^n}{(\omega/{\bar{\omega}}_{Di}-x^2)^2-(\omega_{bi}/{\bar{\omega}}_{Di})^2x^2}dx,\\
&M\left(\frac{\omega}{{\bar\omega}_{Ds}}\right)=-2\frac{\omega}{{\bar\omega}_{Ds}}\Bigg\{\left(1-\frac{\omega_{\ast ni}}{\omega}+\frac{3}{2}\frac{\omega_{\ast Ti}}{\omega}\right)\left[1+\sqrt{\frac{\omega}{{\bar\omega}_{Ds}}}Z\left(\sqrt{\frac{\omega}{{\bar\omega}_{Ds}}}\right)\right]\\
&\qquad \qquad \qquad -\frac{\omega_{\ast Ti}}{\omega}\left[\frac{1}{2}+\frac{\omega}{{\bar\omega}_{Ds}}+\left(\frac{\omega}{{\bar\omega}_{Ds}}\right)^{3/2}Z\left(\sqrt{\frac{\omega}{{\bar\omega}_{Ds}}}\right)\right]\Bigg\},\\
&L\left(\frac{\omega}{{\bar\omega}_{Ds}}\right)=-2\Bigg\{\left(1-\frac{\omega_{\ast ni}}{\omega}+\frac{3}{2}\frac{\omega_{\ast Ti}}{\omega}\right)\left[\frac{1}{2}+\frac{\omega}{{\bar\omega}_{Ds}}+\left(\frac{\omega}{{\bar\omega}_{Ds}}\right)^{3/2}Z\left(\sqrt{\frac{\omega}{{\bar\omega}_{Ds}}}\right)\right]\\
&\qquad \qquad \qquad -\frac{\omega_{\ast Ti}}{\omega}\left[\frac{3}{4}+\frac{1}{2}\frac{\omega}{{\bar\omega}_{Ds}}+\left(\frac{\omega}{{\bar\omega}_{Ds}}\right)^2+\left(\frac{\omega}{{\bar\omega}_{Ds}}\right)^{5/2}Z\left(\sqrt{\frac{\omega}{{\bar\omega}_{Ds}}}\right)\right]\Bigg\}.
\end{aligned}
\end{equation}

Here the magnetic drift orbit precession frequency ${\bar \omega}_{ds}={\bar \omega}_{Ds}m_s\upsilon^2/2T_s$ for deeply trapped particles ($s=i,e$) with ${\bar \omega}_{Ds}=(nq/r)T_s/m_sR_0\omega_{cs}$ and $\omega_{cs}=e_sB/m_sc$; the bounce frequency of deeply trapped ions $\omega_{bi}\equiv (r/R_0)^{1/2}(T_i/m_i)^{1/2}/(qR_0)\approx \epsilon^{1/2}\omega_{ti}$ with $\omega_{ti}=(2T_i/m_i)^{1/2}/qR_0$; and $\tau\equiv T_e/T_i$.

\section*{References}


\begin{thebibliography}{48}%
\makeatletter
\providecommand \@ifxundefined [1]{%
 \@ifx{#1\undefined}
}%
\providecommand \@ifnum [1]{%
 \ifnum #1\expandafter \@firstoftwo
 \else \expandafter \@secondoftwo
 \fi
}%
\providecommand \@ifx [1]{%
 \ifx #1\expandafter \@firstoftwo
 \else \expandafter \@secondoftwo
 \fi
}%
\providecommand \natexlab [1]{#1}%
\providecommand \enquote  [1]{``#1''}%
\providecommand \bibnamefont  [1]{#1}%
\providecommand \bibfnamefont [1]{#1}%
\providecommand \citenamefont [1]{#1}%
\providecommand \href@noop [0]{\@secondoftwo}%
\providecommand \href [0]{\begingroup \@sanitize@url \@href}%
\providecommand \@href[1]{\@@startlink{#1}\@@href}%
\providecommand \@@href[1]{\endgroup#1\@@endlink}%
\providecommand \@sanitize@url [0]{\catcode `\\12\catcode `\$12\catcode
  `\&12\catcode `\#12\catcode `\^12\catcode `\_12\catcode `\%12\relax}%
\providecommand \@@startlink[1]{}%
\providecommand \@@endlink[0]{}%
\providecommand \url  [0]{\begingroup\@sanitize@url \@url }%
\providecommand \@url [1]{\endgroup\@href {#1}{\urlprefix }}%
\providecommand \urlprefix  [0]{URL }%
\providecommand \Eprint [0]{\href }%
\providecommand \doibase [0]{http://dx.doi.org/}%
\providecommand \selectlanguage [0]{\@gobble}%
\providecommand \bibinfo  [0]{\@secondoftwo}%
\providecommand \bibfield  [0]{\@secondoftwo}%
\providecommand \translation [1]{[#1]}%
\providecommand \BibitemOpen [0]{}%
\providecommand \bibitemStop [0]{}%
\providecommand \bibitemNoStop [0]{.\EOS\space}%
\providecommand \EOS [0]{\spacefactor3000\relax}%
\providecommand \BibitemShut  [1]{\csname bibitem#1\endcsname}%
\let\auto@bib@innerbib\@empty
\bibitem [{\citenamefont {Chen}\ and\ \citenamefont {Zonca}(2007)}]{Chen2007}%
  \BibitemOpen
  \bibfield  {author} {\bibinfo {author} {\bibfnamefont {L.}~\bibnamefont
  {Chen}}\ and\ \bibinfo {author} {\bibfnamefont {F.}~\bibnamefont {Zonca}},\
  }\href@noop {} {\bibfield  {journal} {\bibinfo  {journal} {Nucl. Fusion}\
  }\textbf {\bibinfo {volume} {47}},\ \bibinfo {pages} {S727} (\bibinfo {year}
  {2007})}\BibitemShut {NoStop}%
\bibitem [{\citenamefont {Heidbrink}\ \emph {et~al.}(1993)\citenamefont
  {Heidbrink}, \citenamefont {Strait}, \citenamefont {Chu},\ and\ \citenamefont
  {Turnbull}}]{Heidbrink1993}%
  \BibitemOpen
  \bibfield  {author} {\bibinfo {author} {\bibfnamefont {W.~W.}\ \bibnamefont
  {Heidbrink}}, \bibinfo {author} {\bibfnamefont {E.~J.}\ \bibnamefont
  {Strait}}, \bibinfo {author} {\bibfnamefont {M.~S.}\ \bibnamefont {Chu}}, \
  and\ \bibinfo {author} {\bibfnamefont {A.~D.}\ \bibnamefont {Turnbull}},\
  }\href@noop {} {\bibfield  {journal} {\bibinfo  {journal} {Phys. Rev. Lett.}\
  }\textbf {\bibinfo {volume} {71}},\ \bibinfo {pages} {855} (\bibinfo {year}
  {1993})}\BibitemShut {NoStop}%
\bibitem [{\citenamefont {Turnbull}\ \emph {et~al.}(1993)\citenamefont
  {Turnbull}, \citenamefont {Strait}, \citenamefont {Heidbrink}, \citenamefont
  {Chu}, \citenamefont {Duong}, \citenamefont {Greene}, \citenamefont {Lao},
  \citenamefont {Taylor},\ and\ \citenamefont {Thompson}}]{Turnbull1993}%
  \BibitemOpen
  \bibfield  {author} {\bibinfo {author} {\bibfnamefont {A.~D.}\ \bibnamefont
  {Turnbull}}, \bibinfo {author} {\bibfnamefont {E.~J.}\ \bibnamefont
  {Strait}}, \bibinfo {author} {\bibfnamefont {W.~W.}\ \bibnamefont
  {Heidbrink}}, \bibinfo {author} {\bibfnamefont {M.~S.}\ \bibnamefont {Chu}},
  \bibinfo {author} {\bibfnamefont {H.~H.}\ \bibnamefont {Duong}}, \bibinfo
  {author} {\bibfnamefont {J.~W.}\ \bibnamefont {Greene}}, \bibinfo {author}
  {\bibfnamefont {L.~L.}\ \bibnamefont {Lao}}, \bibinfo {author} {\bibfnamefont
  {T.~S.}\ \bibnamefont {Taylor}}, \ and\ \bibinfo {author} {\bibfnamefont
  {S.~J.}\ \bibnamefont {Thompson}},\ }\href@noop {} {\bibfield  {journal}
  {\bibinfo  {journal} {Phys. Fluids B}\ }\textbf {\bibinfo {volume} {5}},\
  \bibinfo {pages} {2546} (\bibinfo {year} {1993})}\BibitemShut {NoStop}%
\bibitem [{\citenamefont {Zonca}\ \emph {et~al.}(1996)\citenamefont {Zonca},
  \citenamefont {Chen},\ and\ \citenamefont {Santoro}}]{Zonca1996}%
  \BibitemOpen
  \bibfield  {author} {\bibinfo {author} {\bibfnamefont {F.}~\bibnamefont
  {Zonca}}, \bibinfo {author} {\bibfnamefont {L.}~\bibnamefont {Chen}}, \ and\
  \bibinfo {author} {\bibfnamefont {R.~A.}\ \bibnamefont {Santoro}},\
  }\href@noop {} {\bibfield  {journal} {\bibinfo  {journal} {Plasma Phys.
  Control. Fusion}\ }\textbf {\bibinfo {volume} {38}},\ \bibinfo {pages} {2011}
  (\bibinfo {year} {1996})}\BibitemShut {NoStop}%
\bibitem [{\citenamefont {Zonca}\ \emph {et~al.}(1999)\citenamefont {Zonca},
  \citenamefont {Chen}, \citenamefont {Dong},\ and\ \citenamefont
  {Santoro}}]{Zonca1999}%
  \BibitemOpen
  \bibfield  {author} {\bibinfo {author} {\bibfnamefont {F.}~\bibnamefont
  {Zonca}}, \bibinfo {author} {\bibfnamefont {L.}~\bibnamefont {Chen}},
  \bibinfo {author} {\bibfnamefont {J.~Q.}\ \bibnamefont {Dong}}, \ and\
  \bibinfo {author} {\bibfnamefont {R.~A.}\ \bibnamefont {Santoro}},\
  }\href@noop {} {\bibfield  {journal} {\bibinfo  {journal} {Phys. Plasmas}\
  }\textbf {\bibinfo {volume} {6}},\ \bibinfo {pages} {1917} (\bibinfo {year}
  {1999})}\BibitemShut {NoStop}%
\bibitem [{\citenamefont {Zonca}\ \emph {et~al.}(2010)\citenamefont {Zonca},
  \citenamefont {Biancalani}, \citenamefont {Chavdarovski}, \citenamefont
  {Chen}, \citenamefont {Troia},\ and\ \citenamefont {Wang}}]{Zonca2010}%
  \BibitemOpen
  \bibfield  {author} {\bibinfo {author} {\bibfnamefont {F.}~\bibnamefont
  {Zonca}}, \bibinfo {author} {\bibfnamefont {A.}~\bibnamefont {Biancalani}},
  \bibinfo {author} {\bibfnamefont {I.}~\bibnamefont {Chavdarovski}}, \bibinfo
  {author} {\bibfnamefont {L.}~\bibnamefont {Chen}}, \bibinfo {author}
  {\bibfnamefont {C.~D.}\ \bibnamefont {Troia}}, \ and\ \bibinfo {author}
  {\bibfnamefont {X.}~\bibnamefont {Wang}},\ }\href@noop {} {\bibfield
  {journal} {\bibinfo  {journal} {Journal of Physics: Conference Series}\
  }\textbf {\bibinfo {volume} {260}},\ \bibinfo {pages} {012022} (\bibinfo
  {year} {2010})}\BibitemShut {NoStop}%
\bibitem [{\citenamefont {Chavdarovski}\ and\ \citenamefont
  {Zonca}(2009)}]{Chavdarovski2009}%
  \BibitemOpen
  \bibfield  {author} {\bibinfo {author} {\bibfnamefont {I.}~\bibnamefont
  {Chavdarovski}}\ and\ \bibinfo {author} {\bibfnamefont {F.}~\bibnamefont
  {Zonca}},\ }\href@noop {} {\bibfield  {journal} {\bibinfo  {journal} {Plasma
  Phys. Control. Fusion}\ }\textbf {\bibinfo {volume} {51}},\ \bibinfo {pages}
  {115001} (\bibinfo {year} {2009})}\BibitemShut {NoStop}%
\bibitem [{\citenamefont {Lauber}\ \emph {et~al.}(2009)\citenamefont {Lauber},
  \citenamefont {Brudgam}, \citenamefont {Curran}, \citenamefont {Igochine},
  \citenamefont {Sassenberg}, \citenamefont {Gunter}, \citenamefont
  {Maraschek}, \citenamefont {Garcia-Munoz}, \citenamefont {Hicks},\ and\
  \citenamefont {the ASDEX Upgrade~Team}}]{Lauber2009}%
  \BibitemOpen
  \bibfield  {author} {\bibinfo {author} {\bibfnamefont {P.}~\bibnamefont
  {Lauber}}, \bibinfo {author} {\bibfnamefont {M.}~\bibnamefont {Brudgam}},
  \bibinfo {author} {\bibfnamefont {D.}~\bibnamefont {Curran}}, \bibinfo
  {author} {\bibfnamefont {V.}~\bibnamefont {Igochine}}, \bibinfo {author}
  {\bibfnamefont {K.}~\bibnamefont {Sassenberg}}, \bibinfo {author}
  {\bibfnamefont {S.}~\bibnamefont {Gunter}}, \bibinfo {author} {\bibfnamefont
  {M.}~\bibnamefont {Maraschek}}, \bibinfo {author} {\bibfnamefont
  {M.}~\bibnamefont {Garcia-Munoz}}, \bibinfo {author} {\bibfnamefont
  {N.}~\bibnamefont {Hicks}}, \ and\ \bibinfo {author} {\bibnamefont {the ASDEX
  Upgrade~Team}},\ }\href@noop {} {\bibfield  {journal} {\bibinfo  {journal}
  {Plasma Phys. Control. Fusion}\ }\textbf {\bibinfo {volume} {51}},\ \bibinfo
  {pages} {124009} (\bibinfo {year} {2009})}\BibitemShut {NoStop}%
\bibitem [{\citenamefont {Chen}\ and\ \citenamefont {Zonca}(2016)}]{Chen2016}%
  \BibitemOpen
  \bibfield  {author} {\bibinfo {author} {\bibfnamefont {L.}~\bibnamefont
  {Chen}}\ and\ \bibinfo {author} {\bibfnamefont {F.}~\bibnamefont {Zonca}},\
  }\href@noop {} {\bibfield  {journal} {\bibinfo  {journal} {Rev. Mod. Phys.}\
  }\textbf {\bibinfo {volume} {88}},\ \bibinfo {pages} {015008} (\bibinfo
  {year} {2016})}\BibitemShut {NoStop}%
\bibitem [{\citenamefont {Zonca}\ \emph {et~al.}(2021)\citenamefont {Zonca},
  \citenamefont {Chen}, \citenamefont {Falessi},\ and\ \citenamefont
  {Qiu}}]{Zonca2021a}%
  \BibitemOpen
  \bibfield  {author} {\bibinfo {author} {\bibfnamefont {F.}~\bibnamefont
  {Zonca}}, \bibinfo {author} {\bibfnamefont {L.}~\bibnamefont {Chen}},
  \bibinfo {author} {\bibfnamefont {M.~V.}\ \bibnamefont {Falessi}}, \ and\
  \bibinfo {author} {\bibfnamefont {Z.}~\bibnamefont {Qiu}},\ }\href@noop {}
  {\bibfield  {journal} {\bibinfo  {journal} {Journal of Physics: Conference
  Series}\ }\textbf {\bibinfo {volume} {1785}},\ \bibinfo {pages} {012005}
  (\bibinfo {year} {2021})}\BibitemShut {NoStop}%
\bibitem [{\citenamefont {Cheng}(1982)}]{Cheng1982}%
  \BibitemOpen
  \bibfield  {author} {\bibinfo {author} {\bibfnamefont {C.}~\bibnamefont
  {Cheng}},\ }\href@noop {} {\bibfield  {journal} {\bibinfo  {journal} {Phys.
  Fluids}\ }\textbf {\bibinfo {volume} {25}},\ \bibinfo {pages} {1020}
  (\bibinfo {year} {1982})}\BibitemShut {NoStop}%
\bibitem [{\citenamefont {Tang}\ \emph {et~al.}(1980)\citenamefont {Tang},
  \citenamefont {Connor},\ and\ \citenamefont {Hastie}}]{Tang1980}%
  \BibitemOpen
  \bibfield  {author} {\bibinfo {author} {\bibfnamefont {W.}~\bibnamefont
  {Tang}}, \bibinfo {author} {\bibfnamefont {J.}~\bibnamefont {Connor}}, \ and\
  \bibinfo {author} {\bibfnamefont {R.}~\bibnamefont {Hastie}},\ }\href@noop {}
  {\bibfield  {journal} {\bibinfo  {journal} {Nucl. Fusion}\ }\textbf {\bibinfo
  {volume} {20(11)}},\ \bibinfo {pages} {1439} (\bibinfo {year}
  {1980})}\BibitemShut {NoStop}%
\bibitem [{\citenamefont {Biglari}\ and\ \citenamefont
  {Chen}(1991)}]{Biglari1991}%
  \BibitemOpen
  \bibfield  {author} {\bibinfo {author} {\bibfnamefont {H.}~\bibnamefont
  {Biglari}}\ and\ \bibinfo {author} {\bibfnamefont {L.}~\bibnamefont {Chen}},\
  }\href@noop {} {\bibfield  {journal} {\bibinfo  {journal} {Phys. Rev. Lett.}\
  }\textbf {\bibinfo {volume} {67}},\ \bibinfo {pages} {3681} (\bibinfo {year}
  {1991})}\BibitemShut {NoStop}%
\bibitem [{\citenamefont {Sharapov}\ \emph {et~al.}(2013)\citenamefont
  {Sharapov}, \citenamefont {Alper}, \citenamefont {Berk}, \citenamefont
  {Borba}, \citenamefont {Breizman}, \citenamefont {Challis}, \citenamefont
  {Classen}, \citenamefont {Edlund}, \citenamefont {Eriksson}, \citenamefont
  {Fasoli}, \citenamefont {Fredrickson}, \citenamefont {Fu}, \citenamefont
  {Garcia-Munoz}, \citenamefont {Gassner}, \citenamefont {Ghantous},
  \citenamefont {Goloborodko}, \citenamefont {Gorelenkov}, \citenamefont
  {Gryaznevich}, \citenamefont {Hacquin}, \citenamefont {Heidbrink},
  \citenamefont {Hellesen}, \citenamefont {Kiptily}, \citenamefont {Kramer},
  \citenamefont {Lauber}, \citenamefont {Lilley}, \citenamefont {Lisak},
  \citenamefont {Nabais}, \citenamefont {Nazikian}, \citenamefont {Nyqvist},
  \citenamefont {Osakabe}, \citenamefont {von Thun}, \citenamefont {Pinches},
  \citenamefont {Podesta}, \citenamefont {Porkolab}, \citenamefont {Shinohara},
  \citenamefont {Schoepf}, \citenamefont {Todo}, \citenamefont {Toi},
  \citenamefont {Zeeland}, \citenamefont {Voitsekhovich}, \citenamefont
  {White}, \citenamefont {Yavorskij}, \citenamefont {TG},\ and\ \citenamefont
  {Contributorsa}}]{Sharapov2013}%
  \BibitemOpen
  \bibfield  {author} {\bibinfo {author} {\bibfnamefont {S.}~\bibnamefont
  {Sharapov}}, \bibinfo {author} {\bibfnamefont {B.}~\bibnamefont {Alper}},
  \bibinfo {author} {\bibfnamefont {H.}~\bibnamefont {Berk}}, \bibinfo {author}
  {\bibfnamefont {D.}~\bibnamefont {Borba}}, \bibinfo {author} {\bibfnamefont
  {B.}~\bibnamefont {Breizman}}, \bibinfo {author} {\bibfnamefont
  {C.}~\bibnamefont {Challis}}, \bibinfo {author} {\bibfnamefont
  {I.}~\bibnamefont {Classen}}, \bibinfo {author} {\bibfnamefont
  {E.}~\bibnamefont {Edlund}}, \bibinfo {author} {\bibfnamefont
  {J.}~\bibnamefont {Eriksson}}, \bibinfo {author} {\bibfnamefont
  {A.}~\bibnamefont {Fasoli}}, \bibinfo {author} {\bibfnamefont
  {E.}~\bibnamefont {Fredrickson}}, \bibinfo {author} {\bibfnamefont
  {G.}~\bibnamefont {Fu}}, \bibinfo {author} {\bibfnamefont {M.}~\bibnamefont
  {Garcia-Munoz}}, \bibinfo {author} {\bibfnamefont {T.}~\bibnamefont
  {Gassner}}, \bibinfo {author} {\bibfnamefont {K.}~\bibnamefont {Ghantous}},
  \bibinfo {author} {\bibfnamefont {V.}~\bibnamefont {Goloborodko}}, \bibinfo
  {author} {\bibfnamefont {N.}~\bibnamefont {Gorelenkov}}, \bibinfo {author}
  {\bibfnamefont {M.}~\bibnamefont {Gryaznevich}}, \bibinfo {author}
  {\bibfnamefont {S.}~\bibnamefont {Hacquin}}, \bibinfo {author} {\bibfnamefont
  {W.}~\bibnamefont {Heidbrink}}, \bibinfo {author} {\bibfnamefont
  {C.}~\bibnamefont {Hellesen}}, \bibinfo {author} {\bibfnamefont
  {V.}~\bibnamefont {Kiptily}}, \bibinfo {author} {\bibfnamefont
  {G.}~\bibnamefont {Kramer}}, \bibinfo {author} {\bibfnamefont
  {P.}~\bibnamefont {Lauber}}, \bibinfo {author} {\bibfnamefont
  {M.}~\bibnamefont {Lilley}}, \bibinfo {author} {\bibfnamefont
  {M.}~\bibnamefont {Lisak}}, \bibinfo {author} {\bibfnamefont
  {F.}~\bibnamefont {Nabais}}, \bibinfo {author} {\bibfnamefont
  {R.}~\bibnamefont {Nazikian}}, \bibinfo {author} {\bibfnamefont
  {R.}~\bibnamefont {Nyqvist}}, \bibinfo {author} {\bibfnamefont
  {M.}~\bibnamefont {Osakabe}}, \bibinfo {author} {\bibfnamefont {C.~P.}\
  \bibnamefont {von Thun}}, \bibinfo {author} {\bibfnamefont {S.}~\bibnamefont
  {Pinches}}, \bibinfo {author} {\bibfnamefont {M.}~\bibnamefont {Podesta}},
  \bibinfo {author} {\bibfnamefont {M.}~\bibnamefont {Porkolab}}, \bibinfo
  {author} {\bibfnamefont {K.}~\bibnamefont {Shinohara}}, \bibinfo {author}
  {\bibfnamefont {K.}~\bibnamefont {Schoepf}}, \bibinfo {author} {\bibfnamefont
  {Y.}~\bibnamefont {Todo}}, \bibinfo {author} {\bibfnamefont {K.}~\bibnamefont
  {Toi}}, \bibinfo {author} {\bibfnamefont {M.~V.}\ \bibnamefont {Zeeland}},
  \bibinfo {author} {\bibfnamefont {I.}~\bibnamefont {Voitsekhovich}}, \bibinfo
  {author} {\bibfnamefont {R.}~\bibnamefont {White}}, \bibinfo {author}
  {\bibfnamefont {V.}~\bibnamefont {Yavorskij}}, \bibinfo {author}
  {\bibfnamefont {I.~E.}\ \bibnamefont {TG}}, \ and\ \bibinfo {author}
  {\bibfnamefont {J.-E.}\ \bibnamefont {Contributorsa}},\ }\href@noop {}
  {\bibfield  {journal} {\bibinfo  {journal} {Nucl. Fusion}\ }\textbf {\bibinfo
  {volume} {53}},\ \bibinfo {pages} {104022} (\bibinfo {year}
  {2013})}\BibitemShut {NoStop}%
\bibitem [{\citenamefont {Gorelenkov}\ \emph {et~al.}(2014)\citenamefont
  {Gorelenkov}, \citenamefont {Pinches},\ and\ \citenamefont
  {Toi}}]{Gorelenkov2014}%
  \BibitemOpen
  \bibfield  {author} {\bibinfo {author} {\bibfnamefont {N.}~\bibnamefont
  {Gorelenkov}}, \bibinfo {author} {\bibfnamefont {S.}~\bibnamefont {Pinches}},
  \ and\ \bibinfo {author} {\bibfnamefont {K.}~\bibnamefont {Toi}},\
  }\href@noop {} {\bibfield  {journal} {\bibinfo  {journal} {Nucl. Fusion}\
  }\textbf {\bibinfo {volume} {54}},\ \bibinfo {pages} {125001} (\bibinfo
  {year} {2014})}\BibitemShut {NoStop}%
\bibitem [{\citenamefont {Heidbrink}\ \emph
  {et~al.}(2021{\natexlab{a}})\citenamefont {Heidbrink}, \citenamefont
  {Zeeland}, \citenamefont {Austin}, \citenamefont {Bierwage}, \citenamefont
  {Chen}, \citenamefont {Choi}, \citenamefont {Lauber}, \citenamefont {Lin},
  \citenamefont {McKee},\ and\ \citenamefont {Spong}}]{Heidbrink2021}%
  \BibitemOpen
  \bibfield  {author} {\bibinfo {author} {\bibfnamefont {W.}~\bibnamefont
  {Heidbrink}}, \bibinfo {author} {\bibfnamefont {M.~V.}\ \bibnamefont
  {Zeeland}}, \bibinfo {author} {\bibfnamefont {M.}~\bibnamefont {Austin}},
  \bibinfo {author} {\bibfnamefont {A.}~\bibnamefont {Bierwage}}, \bibinfo
  {author} {\bibfnamefont {L.}~\bibnamefont {Chen}}, \bibinfo {author}
  {\bibfnamefont {G.}~\bibnamefont {Choi}}, \bibinfo {author} {\bibfnamefont
  {P.}~\bibnamefont {Lauber}}, \bibinfo {author} {\bibfnamefont
  {Z.}~\bibnamefont {Lin}}, \bibinfo {author} {\bibfnamefont {G.}~\bibnamefont
  {McKee}}, \ and\ \bibinfo {author} {\bibfnamefont {D.}~\bibnamefont
  {Spong}},\ }\href@noop {} {\bibfield  {journal} {\bibinfo  {journal} {Nucl.
  Fusion}\ }\textbf {\bibinfo {volume} {61}},\ \bibinfo {pages} {016029}
  (\bibinfo {year} {2021}{\natexlab{a}})}\BibitemShut {NoStop}%
\bibitem [{\citenamefont {Heidbrink}\ \emph
  {et~al.}(2021{\natexlab{b}})\citenamefont {Heidbrink}, \citenamefont
  {Zeeland}, \citenamefont {Austin}, \citenamefont {Crocker}, \citenamefont
  {Du}, \citenamefont {McKee},\ and\ \citenamefont {Spong}}]{Heidbrink2021a}%
  \BibitemOpen
  \bibfield  {author} {\bibinfo {author} {\bibfnamefont {W.}~\bibnamefont
  {Heidbrink}}, \bibinfo {author} {\bibfnamefont {M.~V.}\ \bibnamefont
  {Zeeland}}, \bibinfo {author} {\bibfnamefont {M.}~\bibnamefont {Austin}},
  \bibinfo {author} {\bibfnamefont {N.}~\bibnamefont {Crocker}}, \bibinfo
  {author} {\bibfnamefont {X.}~\bibnamefont {Du}}, \bibinfo {author}
  {\bibfnamefont {G.}~\bibnamefont {McKee}}, \ and\ \bibinfo {author}
  {\bibfnamefont {D.}~\bibnamefont {Spong}},\ }\href@noop {} {\bibfield
  {journal} {\bibinfo  {journal} {Nucl. Fusion}\ }\textbf {\bibinfo {volume}
  {61}},\ \bibinfo {pages} {066031} (\bibinfo {year}
  {2021}{\natexlab{b}})}\BibitemShut {NoStop}%
\bibitem [{\citenamefont {Heidbrink}\ \emph
  {et~al.}(2021{\natexlab{c}})\citenamefont {Heidbrink}, \citenamefont {Choi},
  \citenamefont {Zeeland}, \citenamefont {Austin}, \citenamefont
  {Degrandchamp}, \citenamefont {Spong}, \citenamefont {Bierwage},
  \citenamefont {Crocker}, \citenamefont {Du}, \citenamefont {Lauber},
  \citenamefont {Lin},\ and\ \citenamefont {McKee}}]{Heidbrink2021b}%
  \BibitemOpen
  \bibfield  {author} {\bibinfo {author} {\bibfnamefont {W.}~\bibnamefont
  {Heidbrink}}, \bibinfo {author} {\bibfnamefont {G.}~\bibnamefont {Choi}},
  \bibinfo {author} {\bibfnamefont {M.~V.}\ \bibnamefont {Zeeland}}, \bibinfo
  {author} {\bibfnamefont {M.}~\bibnamefont {Austin}}, \bibinfo {author}
  {\bibfnamefont {G.}~\bibnamefont {Degrandchamp}}, \bibinfo {author}
  {\bibfnamefont {D.}~\bibnamefont {Spong}}, \bibinfo {author} {\bibfnamefont
  {A.}~\bibnamefont {Bierwage}}, \bibinfo {author} {\bibfnamefont
  {N.}~\bibnamefont {Crocker}}, \bibinfo {author} {\bibfnamefont
  {X.}~\bibnamefont {Du}}, \bibinfo {author} {\bibfnamefont {P.}~\bibnamefont
  {Lauber}}, \bibinfo {author} {\bibfnamefont {Z.}~\bibnamefont {Lin}}, \ and\
  \bibinfo {author} {\bibfnamefont {G.}~\bibnamefont {McKee}},\ }\href@noop {}
  {\bibfield  {journal} {\bibinfo  {journal} {Nucl. Fusion}\ }\textbf {\bibinfo
  {volume} {61}},\ \bibinfo {pages} {106021} (\bibinfo {year}
  {2021}{\natexlab{c}})}\BibitemShut {NoStop}%
\bibitem [{\citenamefont {Curran}\ \emph {et~al.}(2012)\citenamefont {Curran},
  \citenamefont {Lauber}, \citenamefont {Carthy}, \citenamefont {da~Graca},
  \citenamefont {Igochine},\ and\ \citenamefont {the ASDEX
  Upgrade~Team}}]{Curran2012}%
  \BibitemOpen
  \bibfield  {author} {\bibinfo {author} {\bibfnamefont {D.}~\bibnamefont
  {Curran}}, \bibinfo {author} {\bibfnamefont {P.}~\bibnamefont {Lauber}},
  \bibinfo {author} {\bibfnamefont {P.~J.~M.}\ \bibnamefont {Carthy}}, \bibinfo
  {author} {\bibfnamefont {S.}~\bibnamefont {da~Graca}}, \bibinfo {author}
  {\bibfnamefont {V.}~\bibnamefont {Igochine}}, \ and\ \bibinfo {author}
  {\bibnamefont {the ASDEX Upgrade~Team}},\ }\href@noop {} {\bibfield
  {journal} {\bibinfo  {journal} {Plasma Phys. Control. Fusion}\ }\textbf
  {\bibinfo {volume} {54}},\ \bibinfo {pages} {055001} (\bibinfo {year}
  {2012})}\BibitemShut {NoStop}%
\bibitem [{\citenamefont {Lauber}(2013)}]{Lauber2013a}%
  \BibitemOpen
  \bibfield  {author} {\bibinfo {author} {\bibfnamefont {P.}~\bibnamefont
  {Lauber}},\ }\href@noop {} {\bibfield  {journal} {\bibinfo  {journal}
  {Physics Reports}\ }\textbf {\bibinfo {volume} {533}},\ \bibinfo {pages} {33}
  (\bibinfo {year} {2013})}\BibitemShut {NoStop}%
\bibitem [{\citenamefont {Chavdarovski}\ and\ \citenamefont
  {Zonca}(2014)}]{Chavdarovski2014}%
  \BibitemOpen
  \bibfield  {author} {\bibinfo {author} {\bibfnamefont {I.}~\bibnamefont
  {Chavdarovski}}\ and\ \bibinfo {author} {\bibfnamefont {F.}~\bibnamefont
  {Zonca}},\ }\href@noop {} {\bibfield  {journal} {\bibinfo  {journal} {Phys.
  Plasmas}\ }\textbf {\bibinfo {volume} {21}},\ \bibinfo {pages} {052506}
  (\bibinfo {year} {2014})}\BibitemShut {NoStop}%
\bibitem [{\citenamefont {Fasoli}\ \emph {et~al.}(2016)\citenamefont {Fasoli},
  \citenamefont {Brunner}, \citenamefont {Cooper}, \citenamefont {Graves},
  \citenamefont {P.Ricci}, \citenamefont {Sauter},\ and\ \citenamefont
  {Villard}}]{Fasoli2016}%
  \BibitemOpen
  \bibfield  {author} {\bibinfo {author} {\bibfnamefont {A.}~\bibnamefont
  {Fasoli}}, \bibinfo {author} {\bibfnamefont {S.}~\bibnamefont {Brunner}},
  \bibinfo {author} {\bibfnamefont {W.}~\bibnamefont {Cooper}}, \bibinfo
  {author} {\bibfnamefont {J.}~\bibnamefont {Graves}}, \bibinfo {author}
  {\bibnamefont {P.Ricci}}, \bibinfo {author} {\bibfnamefont {O.}~\bibnamefont
  {Sauter}}, \ and\ \bibinfo {author} {\bibfnamefont {L.}~\bibnamefont
  {Villard}},\ }\href@noop {} {\bibfield  {journal} {\bibinfo  {journal}
  {Nature Physics}\ }\textbf {\bibinfo {volume} {12}},\ \bibinfo {pages} {411}
  (\bibinfo {year} {2016})}\BibitemShut {NoStop}%
\bibitem [{\citenamefont {Bierwage}\ and\ \citenamefont
  {Lauber}(2017)}]{Bierwage2017}%
  \BibitemOpen
  \bibfield  {author} {\bibinfo {author} {\bibfnamefont {A.}~\bibnamefont
  {Bierwage}}\ and\ \bibinfo {author} {\bibfnamefont {P.}~\bibnamefont
  {Lauber}},\ }\href@noop {} {\bibfield  {journal} {\bibinfo  {journal} {Nucl.
  Fusion}\ }\textbf {\bibinfo {volume} {57}},\ \bibinfo {pages} {116063}
  (\bibinfo {year} {2017})}\BibitemShut {NoStop}%
\bibitem [{\citenamefont {Choi}\ \emph {et~al.}(2021)\citenamefont {Choi},
  \citenamefont {Liu}, \citenamefont {Wei}, \citenamefont {Nicolau},
  \citenamefont {Dong}, \citenamefont {Zhang}, \citenamefont {Lin},
  \citenamefont {Heidbrink},\ and\ \citenamefont {Hahm}}]{Choi2021a}%
  \BibitemOpen
  \bibfield  {author} {\bibinfo {author} {\bibfnamefont {G.}~\bibnamefont
  {Choi}}, \bibinfo {author} {\bibfnamefont {P.}~\bibnamefont {Liu}}, \bibinfo
  {author} {\bibfnamefont {X.}~\bibnamefont {Wei}}, \bibinfo {author}
  {\bibfnamefont {J.}~\bibnamefont {Nicolau}}, \bibinfo {author} {\bibfnamefont
  {G.}~\bibnamefont {Dong}}, \bibinfo {author} {\bibfnamefont {W.}~\bibnamefont
  {Zhang}}, \bibinfo {author} {\bibfnamefont {Z.}~\bibnamefont {Lin}}, \bibinfo
  {author} {\bibfnamefont {W.}~\bibnamefont {Heidbrink}}, \ and\ \bibinfo
  {author} {\bibfnamefont {T.}~\bibnamefont {Hahm}},\ }\href@noop {} {\bibfield
   {journal} {\bibinfo  {journal} {Nucl. Fusion}\ }\textbf {\bibinfo {volume}
  {61}},\ \bibinfo {pages} {066007} (\bibinfo {year} {2021})}\BibitemShut
  {NoStop}%
\bibitem [{\citenamefont {Gorelenkov}\ \emph {et~al.}(2007)\citenamefont
  {Gorelenkov}, \citenamefont {Berk}, \citenamefont {Fredrickson},
  \citenamefont {Sharapov},\ and\ \citenamefont
  {Contributors}}]{Gorelenkov2007}%
  \BibitemOpen
  \bibfield  {author} {\bibinfo {author} {\bibfnamefont {N.}~\bibnamefont
  {Gorelenkov}}, \bibinfo {author} {\bibfnamefont {H.}~\bibnamefont {Berk}},
  \bibinfo {author} {\bibfnamefont {E.}~\bibnamefont {Fredrickson}}, \bibinfo
  {author} {\bibfnamefont {S.}~\bibnamefont {Sharapov}}, \ and\ \bibinfo
  {author} {\bibfnamefont {J.~E.}\ \bibnamefont {Contributors}},\ }\href@noop
  {} {\bibfield  {journal} {\bibinfo  {journal} {Phys. Lett. A}\ }\textbf
  {\bibinfo {volume} {370}},\ \bibinfo {pages} {70} (\bibinfo {year}
  {2007})}\BibitemShut {NoStop}%
\bibitem [{\citenamefont {Gorelenkov}\ \emph {et~al.}(2009)\citenamefont
  {Gorelenkov}, \citenamefont {Zeeland}, \citenamefont {Berk}, \citenamefont
  {Crocker}, \citenamefont {Darrow}, \citenamefont {Fredrickson}, \citenamefont
  {Fu}, \citenamefont {Heidbrink}, \citenamefont {Menard},\ and\ \citenamefont
  {Nazikian}}]{Gorelenkov2009}%
  \BibitemOpen
  \bibfield  {author} {\bibinfo {author} {\bibfnamefont {N.~N.}\ \bibnamefont
  {Gorelenkov}}, \bibinfo {author} {\bibfnamefont {M.~A.~V.}\ \bibnamefont
  {Zeeland}}, \bibinfo {author} {\bibfnamefont {H.~L.}\ \bibnamefont {Berk}},
  \bibinfo {author} {\bibfnamefont {N.~A.}\ \bibnamefont {Crocker}}, \bibinfo
  {author} {\bibfnamefont {D.}~\bibnamefont {Darrow}}, \bibinfo {author}
  {\bibfnamefont {E.}~\bibnamefont {Fredrickson}}, \bibinfo {author}
  {\bibfnamefont {G.-Y.}\ \bibnamefont {Fu}}, \bibinfo {author} {\bibfnamefont
  {W.~W.}\ \bibnamefont {Heidbrink}}, \bibinfo {author} {\bibfnamefont
  {J.}~\bibnamefont {Menard}}, \ and\ \bibinfo {author} {\bibfnamefont
  {R.}~\bibnamefont {Nazikian}},\ }\href@noop {} {\bibfield  {journal}
  {\bibinfo  {journal} {Phys. Plasmas}\ }\textbf {\bibinfo {volume} {16}},\
  \bibinfo {pages} {056107} (\bibinfo {year} {2009})}\BibitemShut {NoStop}%
\bibitem [{\citenamefont {Chen}\ and\ \citenamefont {Zonca}(2017)}]{Chen2017}%
  \BibitemOpen
  \bibfield  {author} {\bibinfo {author} {\bibfnamefont {L.}~\bibnamefont
  {Chen}}\ and\ \bibinfo {author} {\bibfnamefont {F.}~\bibnamefont {Zonca}},\
  }\href@noop {} {\bibfield  {journal} {\bibinfo  {journal} {Phys. Plasmas}\
  }\textbf {\bibinfo {volume} {24}},\ \bibinfo {pages} {072511} (\bibinfo
  {year} {2017})}\BibitemShut {NoStop}%
\bibitem [{\citenamefont {Zonca}\ and\ \citenamefont
  {Chen}(2014{\natexlab{a}})}]{Zonca2014}%
  \BibitemOpen
  \bibfield  {author} {\bibinfo {author} {\bibfnamefont {F.}~\bibnamefont
  {Zonca}}\ and\ \bibinfo {author} {\bibfnamefont {L.}~\bibnamefont {Chen}},\
  }\href@noop {} {\bibfield  {journal} {\bibinfo  {journal} {Phys. Plasmas}\
  }\textbf {\bibinfo {volume} {21}},\ \bibinfo {pages} {072120} (\bibinfo
  {year} {2014}{\natexlab{a}})}\BibitemShut {NoStop}%
\bibitem [{\citenamefont {Zonca}\ and\ \citenamefont
  {Chen}(2014{\natexlab{b}})}]{Zonca2014a}%
  \BibitemOpen
  \bibfield  {author} {\bibinfo {author} {\bibfnamefont {F.}~\bibnamefont
  {Zonca}}\ and\ \bibinfo {author} {\bibfnamefont {L.}~\bibnamefont {Chen}},\
  }\href@noop {} {\bibfield  {journal} {\bibinfo  {journal} {Phys. Plasmas}\
  }\textbf {\bibinfo {volume} {21}},\ \bibinfo {pages} {072121} (\bibinfo
  {year} {2014}{\natexlab{b}})}\BibitemShut {NoStop}%
\bibitem [{\citenamefont {Varela}\ \emph {et~al.}(2018)\citenamefont {Varela},
  \citenamefont {Spong}, \citenamefont {Garcia}, \citenamefont {Huang},
  \citenamefont {Murakami}, \citenamefont {Garofalo}, \citenamefont {Qian},
  \citenamefont {Holcomb}, \citenamefont {Hyatt}, \citenamefont {Ferron},
  \citenamefont {Collins}, \citenamefont {Ren}, \citenamefont {McClenaghan},\
  and\ \citenamefont {Guo}}]{Varela2018}%
  \BibitemOpen
  \bibfield  {author} {\bibinfo {author} {\bibfnamefont {J.}~\bibnamefont
  {Varela}}, \bibinfo {author} {\bibfnamefont {D.}~\bibnamefont {Spong}},
  \bibinfo {author} {\bibfnamefont {L.}~\bibnamefont {Garcia}}, \bibinfo
  {author} {\bibfnamefont {J.}~\bibnamefont {Huang}}, \bibinfo {author}
  {\bibfnamefont {M.}~\bibnamefont {Murakami}}, \bibinfo {author}
  {\bibfnamefont {A.}~\bibnamefont {Garofalo}}, \bibinfo {author}
  {\bibfnamefont {J.}~\bibnamefont {Qian}}, \bibinfo {author} {\bibfnamefont
  {C.}~\bibnamefont {Holcomb}}, \bibinfo {author} {\bibfnamefont
  {A.}~\bibnamefont {Hyatt}}, \bibinfo {author} {\bibfnamefont
  {J.}~\bibnamefont {Ferron}}, \bibinfo {author} {\bibfnamefont
  {C.}~\bibnamefont {Collins}}, \bibinfo {author} {\bibfnamefont
  {Q.}~\bibnamefont {Ren}}, \bibinfo {author} {\bibfnamefont {J.}~\bibnamefont
  {McClenaghan}}, \ and\ \bibinfo {author} {\bibfnamefont {W.}~\bibnamefont
  {Guo}},\ }\href@noop {} {\bibfield  {journal} {\bibinfo  {journal} {Nucl.
  Fusion}\ }\textbf {\bibinfo {volume} {58}},\ \bibinfo {pages} {076017}
  (\bibinfo {year} {2018})}\BibitemShut {NoStop}%
\bibitem [{\citenamefont {Chen}\ \emph {et~al.}(1984)\citenamefont {Chen},
  \citenamefont {White},\ and\ \citenamefont {Rosenbluth}}]{Chen1984}%
  \BibitemOpen
  \bibfield  {author} {\bibinfo {author} {\bibfnamefont {L.}~\bibnamefont
  {Chen}}, \bibinfo {author} {\bibfnamefont {R.~B.}\ \bibnamefont {White}}, \
  and\ \bibinfo {author} {\bibfnamefont {M.}~\bibnamefont {Rosenbluth}},\
  }\href@noop {} {\bibfield  {journal} {\bibinfo  {journal} {Phys. Rev. Lett.}\
  }\textbf {\bibinfo {volume} {52}},\ \bibinfo {pages} {1122} (\bibinfo {year}
  {1984})}\BibitemShut {NoStop}%
\bibitem [{\citenamefont {Chen}(1994)}]{Chen1994}%
  \BibitemOpen
  \bibfield  {author} {\bibinfo {author} {\bibfnamefont {L.}~\bibnamefont
  {Chen}},\ }\href@noop {} {\bibfield  {journal} {\bibinfo  {journal} {Phys.
  Plasmas}\ }\textbf {\bibinfo {volume} {1}},\ \bibinfo {pages} {1519}
  (\bibinfo {year} {1994})}\BibitemShut {NoStop}%
\bibitem [{\citenamefont {Tsai}\ and\ \citenamefont {Chen}(1993)}]{Tsai1993}%
  \BibitemOpen
  \bibfield  {author} {\bibinfo {author} {\bibfnamefont {S.}~\bibnamefont
  {Tsai}}\ and\ \bibinfo {author} {\bibfnamefont {L.}~\bibnamefont {Chen}},\
  }\href@noop {} {\bibfield  {journal} {\bibinfo  {journal} {Phys. Fluids B}\
  }\textbf {\bibinfo {volume} {5}},\ \bibinfo {pages} {3284} (\bibinfo {year}
  {1993})}\BibitemShut {NoStop}%
\bibitem [{\citenamefont {Zonca}\ and\ \citenamefont {Chen}(2006)}]{Zonca2006}%
  \BibitemOpen
  \bibfield  {author} {\bibinfo {author} {\bibfnamefont {F.}~\bibnamefont
  {Zonca}}\ and\ \bibinfo {author} {\bibfnamefont {L.}~\bibnamefont {Chen}},\
  }\href@noop {} {\bibfield  {journal} {\bibinfo  {journal} {Plasma Phys.
  Control. Fusion}\ }\textbf {\bibinfo {volume} {48}},\ \bibinfo {pages} {537}
  (\bibinfo {year} {2006})}\BibitemShut {NoStop}%
\bibitem [{\citenamefont {Zonca}\ \emph {et~al.}(2007)\citenamefont {Zonca},
  \citenamefont {Buratti}, \citenamefont {Cardinali}, \citenamefont {Chen},
  \citenamefont {Dong}, \citenamefont {Long}, \citenamefont {Milovanov},
  \citenamefont {Romanelli}, \citenamefont {Smeulders}, \citenamefont {Wang},
  \citenamefont {Wang}, \citenamefont {Castaldo}, \citenamefont {Cesario},
  \citenamefont {Giovannozzi}, \citenamefont {Marinucci},\ and\ \citenamefont
  {Ridolfini}}]{Zonca2007}%
  \BibitemOpen
  \bibfield  {author} {\bibinfo {author} {\bibfnamefont {F.}~\bibnamefont
  {Zonca}}, \bibinfo {author} {\bibfnamefont {P.}~\bibnamefont {Buratti}},
  \bibinfo {author} {\bibfnamefont {A.}~\bibnamefont {Cardinali}}, \bibinfo
  {author} {\bibfnamefont {L.}~\bibnamefont {Chen}}, \bibinfo {author}
  {\bibfnamefont {J.-Q.}\ \bibnamefont {Dong}}, \bibinfo {author}
  {\bibfnamefont {Y.-X.}\ \bibnamefont {Long}}, \bibinfo {author}
  {\bibfnamefont {A.}~\bibnamefont {Milovanov}}, \bibinfo {author}
  {\bibfnamefont {F.}~\bibnamefont {Romanelli}}, \bibinfo {author}
  {\bibfnamefont {P.}~\bibnamefont {Smeulders}}, \bibinfo {author}
  {\bibfnamefont {L.}~\bibnamefont {Wang}}, \bibinfo {author} {\bibfnamefont
  {Z.-T.}\ \bibnamefont {Wang}}, \bibinfo {author} {\bibfnamefont
  {C.}~\bibnamefont {Castaldo}}, \bibinfo {author} {\bibfnamefont
  {R.}~\bibnamefont {Cesario}}, \bibinfo {author} {\bibfnamefont
  {E.}~\bibnamefont {Giovannozzi}}, \bibinfo {author} {\bibfnamefont
  {M.}~\bibnamefont {Marinucci}}, \ and\ \bibinfo {author} {\bibfnamefont
  {V.~P.}\ \bibnamefont {Ridolfini}},\ }\href@noop {} {\bibfield  {journal}
  {\bibinfo  {journal} {Nucl. Fusion}\ }\textbf {\bibinfo {volume} {47}},\
  \bibinfo {pages} {1588} (\bibinfo {year} {2007})}\BibitemShut {NoStop}%
\bibitem [{\citenamefont {Ma}\ \emph {et~al.}(2022)\citenamefont {Ma},
  \citenamefont {Chen}, \citenamefont {Zonca}, \citenamefont {Li},\ and\
  \citenamefont {Qiu}}]{Ma2022}%
  \BibitemOpen
  \bibfield  {author} {\bibinfo {author} {\bibfnamefont {R.}~\bibnamefont
  {Ma}}, \bibinfo {author} {\bibfnamefont {L.}~\bibnamefont {Chen}}, \bibinfo
  {author} {\bibfnamefont {F.}~\bibnamefont {Zonca}}, \bibinfo {author}
  {\bibfnamefont {Y.}~\bibnamefont {Li}}, \ and\ \bibinfo {author}
  {\bibfnamefont {Z.}~\bibnamefont {Qiu}},\ }\href@noop {} {\bibfield
  {journal} {\bibinfo  {journal} {Plasma Phys. Control. Fusion}\ }\textbf
  {\bibinfo {volume} {64}},\ \bibinfo {pages} {035019} (\bibinfo {year}
  {2022})}\BibitemShut {NoStop}%
\bibitem [{\citenamefont {Lao}\ \emph {et~al.}(1985)\citenamefont {Lao},
  \citenamefont {John},\ and\ \citenamefont {Stambaugh}}]{Lao1985}%
  \BibitemOpen
  \bibfield  {author} {\bibinfo {author} {\bibfnamefont {L.~L.}\ \bibnamefont
  {Lao}}, \bibinfo {author} {\bibfnamefont {H.~S.}\ \bibnamefont {John}}, \
  and\ \bibinfo {author} {\bibfnamefont {R.~D.}\ \bibnamefont {Stambaugh}},\
  }\href@noop {} {\bibfield  {journal} {\bibinfo  {journal} {Nucl. Fusion}\
  }\textbf {\bibinfo {volume} {25}},\ \bibinfo {pages} {1611} (\bibinfo {year}
  {1985})}\BibitemShut {NoStop}%
\bibitem [{\citenamefont {Pankin}\ \emph {et~al.}(2004)\citenamefont {Pankin},
  \citenamefont {McCune}, \citenamefont {Andre}, \citenamefont {Bateman},\ and\
  \citenamefont {Kritz}}]{Pankin2004}%
  \BibitemOpen
  \bibfield  {author} {\bibinfo {author} {\bibfnamefont {A.}~\bibnamefont
  {Pankin}}, \bibinfo {author} {\bibfnamefont {D.}~\bibnamefont {McCune}},
  \bibinfo {author} {\bibfnamefont {R.}~\bibnamefont {Andre}}, \bibinfo
  {author} {\bibfnamefont {G.}~\bibnamefont {Bateman}}, \ and\ \bibinfo
  {author} {\bibfnamefont {A.}~\bibnamefont {Kritz}},\ }\href@noop {}
  {\bibfield  {journal} {\bibinfo  {journal} {Comput. Phys. Commun.}\ }\textbf
  {\bibinfo {volume} {159}},\ \bibinfo {pages} {157} (\bibinfo {year}
  {2004})}\BibitemShut {NoStop}%
\bibitem [{\citenamefont {Zonca}\ and\ \citenamefont {Chen}(2000)}]{Zonca2000}%
  \BibitemOpen
  \bibfield  {author} {\bibinfo {author} {\bibfnamefont {F.}~\bibnamefont
  {Zonca}}\ and\ \bibinfo {author} {\bibfnamefont {L.}~\bibnamefont {Chen}},\
  }\href@noop {} {\bibfield  {journal} {\bibinfo  {journal} {Phys. Plasmas}\
  }\textbf {\bibinfo {volume} {7}},\ \bibinfo {pages} {4600} (\bibinfo {year}
  {2000})}\BibitemShut {NoStop}%
\bibitem [{\citenamefont {Zonca}\ \emph {et~al.}(2002)\citenamefont {Zonca},
  \citenamefont {Briguglio}, \citenamefont {Chen}, \citenamefont {Dettrick},
  \citenamefont {Fogaccia}, \citenamefont {Testa},\ and\ \citenamefont
  {Vlad}}]{Zonca2002}%
  \BibitemOpen
  \bibfield  {author} {\bibinfo {author} {\bibfnamefont {F.}~\bibnamefont
  {Zonca}}, \bibinfo {author} {\bibfnamefont {S.}~\bibnamefont {Briguglio}},
  \bibinfo {author} {\bibfnamefont {L.}~\bibnamefont {Chen}}, \bibinfo {author}
  {\bibfnamefont {S.}~\bibnamefont {Dettrick}}, \bibinfo {author}
  {\bibfnamefont {G.}~\bibnamefont {Fogaccia}}, \bibinfo {author}
  {\bibfnamefont {D.}~\bibnamefont {Testa}}, \ and\ \bibinfo {author}
  {\bibfnamefont {G.}~\bibnamefont {Vlad}},\ }\href@noop {} {\bibfield
  {journal} {\bibinfo  {journal} {Phys. Plasmas}\ }\textbf {\bibinfo {volume}
  {9(12)}},\ \bibinfo {pages} {4939} (\bibinfo {year} {2002})}\BibitemShut
  {NoStop}%
\bibitem [{\citenamefont {Connor}\ \emph {et~al.}(1978)\citenamefont {Connor},
  \citenamefont {Hastie},\ and\ \citenamefont {Taylor}}]{Connor1978}%
  \BibitemOpen
  \bibfield  {author} {\bibinfo {author} {\bibfnamefont {J.~W.}\ \bibnamefont
  {Connor}}, \bibinfo {author} {\bibfnamefont {R.~J.}\ \bibnamefont {Hastie}},
  \ and\ \bibinfo {author} {\bibfnamefont {J.~B.}\ \bibnamefont {Taylor}},\
  }\href@noop {} {\bibfield  {journal} {\bibinfo  {journal} {Phys. Rev. Lett.}\
  }\textbf {\bibinfo {volume} {40}},\ \bibinfo {pages} {396} (\bibinfo {year}
  {1978})}\BibitemShut {NoStop}%
\bibitem [{\citenamefont {Zonca}\ \emph {et~al.}(2009)\citenamefont {Zonca},
  \citenamefont {Chen}, \citenamefont {Botrugno}, \citenamefont {Buratti},
  \citenamefont {Cardinali}, \citenamefont {Cesario}, \citenamefont
  {Ridolfini},\ and\ \citenamefont {the JET-EFDA~contributors}}]{Zonca2009}%
  \BibitemOpen
  \bibfield  {author} {\bibinfo {author} {\bibfnamefont {F.}~\bibnamefont
  {Zonca}}, \bibinfo {author} {\bibfnamefont {L.}~\bibnamefont {Chen}},
  \bibinfo {author} {\bibfnamefont {A.}~\bibnamefont {Botrugno}}, \bibinfo
  {author} {\bibfnamefont {P.}~\bibnamefont {Buratti}}, \bibinfo {author}
  {\bibfnamefont {A.}~\bibnamefont {Cardinali}}, \bibinfo {author}
  {\bibfnamefont {R.}~\bibnamefont {Cesario}}, \bibinfo {author} {\bibfnamefont
  {V.~P.}\ \bibnamefont {Ridolfini}}, \ and\ \bibinfo {author} {\bibnamefont
  {the JET-EFDA~contributors}},\ }\href@noop {} {\bibfield  {journal} {\bibinfo
   {journal} {Nucl. Fusion}\ }\textbf {\bibinfo {volume} {49}},\ \bibinfo
  {pages} {085009} (\bibinfo {year} {2009})}\BibitemShut {NoStop}%
\bibitem [{\citenamefont {Chen}(2020)}]{Chen2020}%
  \BibitemOpen
  \bibfield  {author} {\bibinfo {author} {\bibfnamefont {L.}~\bibnamefont
  {Chen}},\ }\href@noop {} {\bibfield  {journal} {\bibinfo  {journal} {private
  notes}\ } (\bibinfo {year} {2020})}\BibitemShut {NoStop}%
\bibitem [{\citenamefont {Rosenbluth}\ and\ \citenamefont
  {Hinton}(1998)}]{Rosenbluth1998}%
  \BibitemOpen
  \bibfield  {author} {\bibinfo {author} {\bibfnamefont {M.~N.}\ \bibnamefont
  {Rosenbluth}}\ and\ \bibinfo {author} {\bibfnamefont {F.~L.}\ \bibnamefont
  {Hinton}},\ }\href@noop {} {\bibfield  {journal} {\bibinfo  {journal} {Phys.
  Rev. Lett.}\ }\textbf {\bibinfo {volume} {80}},\ \bibinfo {pages} {724}
  (\bibinfo {year} {1998})}\BibitemShut {NoStop}%
\bibitem [{\citenamefont {Graves}\ \emph {et~al.}(2000)\citenamefont {Graves},
  \citenamefont {Hastie},\ and\ \citenamefont {Hopcraft}}]{Graves2000}%
  \BibitemOpen
  \bibfield  {author} {\bibinfo {author} {\bibfnamefont {J.~P.}\ \bibnamefont
  {Graves}}, \bibinfo {author} {\bibfnamefont {R.~J.}\ \bibnamefont {Hastie}},
  \ and\ \bibinfo {author} {\bibfnamefont {K.~I.}\ \bibnamefont {Hopcraft}},\
  }\href@noop {} {\bibfield  {journal} {\bibinfo  {journal} {Plasma Phys.
  Control. Fusion}\ }\textbf {\bibinfo {volume} {42}},\ \bibinfo {pages} {1049}
  (\bibinfo {year} {2000})}\BibitemShut {NoStop}%
\bibitem [{\citenamefont {Falessi}\ \emph {et~al.}(2019)\citenamefont
  {Falessi}, \citenamefont {Carlevaro}, \citenamefont {Fusco}, \citenamefont
  {Vlad},\ and\ \citenamefont {Zonca}}]{Falessi2019a}%
  \BibitemOpen
  \bibfield  {author} {\bibinfo {author} {\bibfnamefont {M.~V.}\ \bibnamefont
  {Falessi}}, \bibinfo {author} {\bibfnamefont {N.}~\bibnamefont {Carlevaro}},
  \bibinfo {author} {\bibfnamefont {V.}~\bibnamefont {Fusco}}, \bibinfo
  {author} {\bibfnamefont {G.}~\bibnamefont {Vlad}}, \ and\ \bibinfo {author}
  {\bibfnamefont {F.}~\bibnamefont {Zonca}},\ }\href@noop {} {\bibfield
  {journal} {\bibinfo  {journal} {Phys. Plasmas}\ }\textbf {\bibinfo {volume}
  {26}},\ \bibinfo {pages} {082502} (\bibinfo {year} {2019})}\BibitemShut
  {NoStop}%
\bibitem [{\citenamefont {Falessi}\ \emph {et~al.}(2020)\citenamefont
  {Falessi}, \citenamefont {Carlevaro}, \citenamefont {Fusco}, \citenamefont
  {Giovannozzi}, \citenamefont {Lauber}, \citenamefont {Vlad},\ and\
  \citenamefont {Zonca}}]{Falessi2020}%
  \BibitemOpen
  \bibfield  {author} {\bibinfo {author} {\bibfnamefont {M.~V.}\ \bibnamefont
  {Falessi}}, \bibinfo {author} {\bibfnamefont {N.}~\bibnamefont {Carlevaro}},
  \bibinfo {author} {\bibfnamefont {V.}~\bibnamefont {Fusco}}, \bibinfo
  {author} {\bibfnamefont {E.}~\bibnamefont {Giovannozzi}}, \bibinfo {author}
  {\bibfnamefont {P.}~\bibnamefont {Lauber}}, \bibinfo {author} {\bibfnamefont
  {G.}~\bibnamefont {Vlad}}, \ and\ \bibinfo {author} {\bibfnamefont
  {F.}~\bibnamefont {Zonca}},\ }\href@noop {} {\bibfield  {journal} {\bibinfo
  {journal} {J. Plasma Phys.}\ }\textbf {\bibinfo {volume} {86}},\ \bibinfo
  {pages} {845860501} (\bibinfo {year} {2020})}\BibitemShut {NoStop}%
\bibitem [{\citenamefont {Chen}\ \emph {et~al.}(2016)\citenamefont {Chen},
  \citenamefont {Ding}, \citenamefont {Brower}, \citenamefont {Finkenthal},
  \citenamefont {Muscatello}, \citenamefont {Taussig},\ and\ \citenamefont
  {Boivin}}]{Chen2016b}%
  \BibitemOpen
  \bibfield  {author} {\bibinfo {author} {\bibfnamefont {J.}~\bibnamefont
  {Chen}}, \bibinfo {author} {\bibfnamefont {W.~X.}\ \bibnamefont {Ding}},
  \bibinfo {author} {\bibfnamefont {D.~L.}\ \bibnamefont {Brower}}, \bibinfo
  {author} {\bibfnamefont {D.}~\bibnamefont {Finkenthal}}, \bibinfo {author}
  {\bibfnamefont {C.}~\bibnamefont {Muscatello}}, \bibinfo {author}
  {\bibfnamefont {D.}~\bibnamefont {Taussig}}, \ and\ \bibinfo {author}
  {\bibfnamefont {R.}~\bibnamefont {Boivin}},\ }\href@noop {} {\bibfield
  {journal} {\bibinfo  {journal} {Review of Scientific Instruments}\ }\textbf
  {\bibinfo {volume} {87}},\ \bibinfo {pages} {11E108} (\bibinfo {year}
  {2016})}\BibitemShut {NoStop}%
\end{thebibliography}
\end{document}